\documentclass[aps,prd,amsmath,amssymb,10pt,letterpaper,balancelastpage,superscriptaddress,nofootinbib,floatfix,notitlepage,twocolumn]{revtex4-2}

\usepackage{graphicx}	
\usepackage{subcaption}
\usepackage{booktabs}
\usepackage{ragged2e}	
\usepackage{bm}		    
\usepackage[sort&compress]{natbib}	 	
	\setcitestyle{square,numbers,comma}	
\usepackage[colorlinks=true,urlcolor=blue,linkcolor=black,citecolor=red]{hyperref}	

\usepackage{xcolor}
\usepackage{fontawesome5}
\definecolor{orcidlogocol}{rgb}{0.65, 0.807, 0.223}
\newcommand{\orcid}[1]{\,\href{https://orcid.org/#1}{\textcolor{orcidlogocol}{\footnotesize\faOrcid}}\,}
\renewcommand{\eqref}[2][]{Eq{#1}.~(\ref{#2})}		

\begin{document}
\title{Stable cosmology under flexible covariances: \texorpdfstring{\\}{ }Rescaling and cross-survey correlations in DESI BAO and supernova data}

\author{Eleni-Anna Kolonia\orcid{0000-0002-2695-8781}}
\email{ekolonia@uwaterloo.ca}
\affiliation{Perimeter Institute for Theoretical Physics, 31 Caroline Street North, Waterloo, Ontario, N2L 2Y5, Canada}
\affiliation{Department of Physics and Astronomy, University of Waterloo, 200 University Avenue West, Waterloo, Ontario, N2L 3G1, Canada}
\affiliation{Physics Department, Swansea University, SA2 8PP, UK}

\author{Haoye Li}
\affiliation{Waterloo Centre for Astrophysics, University of Waterloo, 200 University Avenue West, Waterloo, Ontario, N2L 3G1, Canada}
\affiliation{Department of Physics and Astronomy, University of Waterloo, 200 University Avenue West, Waterloo, Ontario, N2L 3G1, Canada}
\affiliation{Perimeter Institute for Theoretical Physics, 31 Caroline Street North, Waterloo, Ontario, N2L 2Y5, Canada}

\author{Hanyu Zhang\orcid{0000-0001-6847-5254}}
\affiliation{Waterloo Centre for Astrophysics, University of Waterloo, 200 University Avenue West, Waterloo, Ontario, N2L 3G1, Canada}
\affiliation{Department of Physics and Astronomy, University of Waterloo, 200 University Avenue West, Waterloo, Ontario, N2L 3G1, Canada}

\author{Niayesh Afshordi\orcid{0000-0002-9940-7040}}
\affiliation{Waterloo Centre for Astrophysics, University of Waterloo, 200 University Avenue West, Waterloo, Ontario, N2L 3G1, Canada}
\affiliation{Department of Physics and Astronomy, University of Waterloo, 200 University Avenue West, Waterloo, Ontario, N2L 3G1, Canada}
\affiliation{Perimeter Institute for Theoretical Physics, 31 Caroline Street North, Waterloo, Ontario, N2L 2Y5, Canada}

\begin{abstract}
Combinations of cosmic microwave background (CMB), baryon acoustic oscillation (BAO), and Type Ia supernova (SN) data currently favor evolving dark energy over $\Lambda$CDM at the $\sim3$--$4\sigma$ level, making it important to test whether this preference could be driven by imperfectly modeled uncertainties. We introduce a phenomenological framework in which the published covariance of each dataset is rescaled by a free parameter and the SN compilations are allowed effective cross-survey correlations. We apply it to DESI DR2 BAO, Planck 2018 temperature and polarization data, and a joint likelihood of Pantheon+, Union3, and DES--Dovekie constructed on a common redshift basis. The DESI, Union3, and DES--Dovekie data are consistent with their published covariances, while Pantheon+ shows a mild preference for additional covariance. In all cases the data still allow error bars up to $\sim1.7$ times larger than published at 95\% credibility, yet freeing the rescaling parameters improves the best fit by $\Delta\chi^2<1$. Pantheon+ and Union3 exhibit a clear positive effective correlation, $r\simeq0.4$--$0.5$, plausibly reflecting their shared supernovae; the Pantheon+--DES--Dovekie correlation is weaker and model dependent, and the Union3--DES--Dovekie correlation is consistent with zero. The added covariance freedom broadens the cosmological constraints without shifting them: for the full data combination, the improvement of $w_0w_a$CDM over $\Lambda$CDM changes only from $\Delta\chi^2=12.8$ to $12.3$ (nominally $3.15\sigma$ to $3.07\sigma$). The preference for evolving dark energy is therefore robust to covariance misestimation of the form considered here.

\end{abstract}

\maketitle

\section{Introduction}
\label{sec:introduction}
Precision cosmology relies on multiple independent probes of the expansion history of the Universe, most notably the cosmic microwave background (CMB), baryon acoustic oscillations (BAO), and Type Ia supernovae (SN). Together, these observations provide complementary constraints on both early- and late-time cosmology, and have played a central role in establishing the standard $\Lambda$CDM model \cite{Planck2018Parameters,DESI2025DR2BAO, Eisenstein1998Complementarity,Seo2003BAO,Blake2003BAO,Hu2002CMB,Riess1998,Perlmutter1999}.

The CMB provides a snapshot of the early Universe and tightly constrains the primordial cosmological parameters, while BAO measurements trace the imprint of acoustic oscillations in the late-time matter distribution, serving as a standard ruler for distance measurements. Supernovae, on the other hand, probe the expansion history at low to intermediate redshifts through standardized luminosity distances, making them particularly sensitive to late-time acceleration.

The combination of these datasets has led to an overall remarkably consistent cosmological picture. However, the increasing precision of current surveys has also made it possible to probe potential tensions between different measurements. In particular, recent BAO results from the Dark Energy Spectroscopic Instrument (DESI) have enabled high-precision distance measurements over a wide redshift range, allowing for detailed comparisons with previous surveys such as SDSS \cite{DESI2013WhitePaper,DESI2016InstrumentDesign,DESI2022Instrumentation,Blanton2017SDSSIV,Dawson2013BOSS,Dawson2016eBOSS,Eisenstein2011SDSSIII}.

While early comparisons between DESI DR1 and SDSS indicated mild discrepancies in certain redshift bins, with some measurements showing non-overlapping uncertainties, the subsequent DESI DR2 release suggests improved consistency between the datasets \cite{DESI2024DR1,DESI2025DR2BAO,eBOSS2021}. Motivated by these comparisons, we test whether allowing additional covariance freedom in the DESI DR2 BAO likelihood materially changes the cosmological constraints obtained in combination with Planck CMB data.

A similar situation arises in the case of Type Ia supernova datasets. Modern compilations such as Pantheon+, Union3, and DES combine observations obtained with different calibration procedures, selection functions, and light-curve fitting methodologies \cite{PantheonPlusData,Union3,DES5YR,DESDovekie}.  Because some supernovae are common to multiple datasets, these samples must be merged with caution, although recent advances have been achieved in this direction \cite{Camilleri2026Unite}. 

These issues point to a broader methodological challenge in cosmological analyses: systematic uncertainties are often treated in a dataset-dependent manner, and correlations between different surveys are either neglected or only approximately modeled. As a result, combining heterogeneous datasets within a single likelihood framework may lead to biased constraints or underestimated uncertainties.

What brings urgency to this question is that different combinations of these datasets with CMB appear to disfavor the $\Lambda$CDM cosmological model, in favor of a dynamical dark energy, often quantified using a Chevallier--Polarski--Linder (CPL) parametrization \cite{ChevallierPolarski2001,Linder2003}, often referred to as $w_0w_a$CDM model. Depending on the combination, the tension with $\Lambda$CDM is about 3-4$\sigma$ \cite{DESI2024DR1,DESI2025DR2BAO}, although the exact level of discrepancy depends on the choice of beyond $\Lambda$CDM model (e.g., \cite{2025PhRvD.112h3511L,2025NatAs...9.1879G,2025JCAP...09..053I, 2025JCAP...08..014C,2025PhRvD.111l3511S,2025PhRvD.112d3513S,2025PhRvD.112f3551G, 2025PhRvD.112f3548C, 2026PhRvD.113b3515P, 2026PhRvD.113d1304P,Wen:2026xlj}
). Yet, the question remains as to whether the datasets are consistent or that the tension may be driven by internal systematics.   
In this work, we develop a phenomenological framework to incorporate such residual systematics and correlations in a unified way, in the context of a reference model. For the BAO measurements, we introduce a covariance-rescaling parameter that allows additional covariance freedom relative to the published covariance matrix. For supernova datasets, we extend this idea by introducing survey-dependent systematics and inter-survey correlation parameters. To enable a consistent joint analysis, we further rebin different supernova compilations onto a common redshift grid.
This framework allows us to assess the sensitivity of the inferred cosmological constraints to additional covariance freedom and phenomenological inter-survey correlations.

The remainder of this paper is organized as follows: Section~\ref{sec:method} presents the methodological framework, Section~\ref{sec:data} describes the datasets, Section~\ref{sec:bao} and Section~\ref{sec:sn} present the BAO and SN results respectively, and Section~\ref{sec:conclusions} summarizes our findings.

\section{Methodology} 
\label{sec:method}

In this section, we discuss our methodology to measure a phenomenological covariance matrix, through comparison of diverse correlated datasets, with potentially unknown systematic errors, with predictions of specific cosmological models. 

We focus our analysis on the standard flat $\Lambda$CDM, as well as dynamical dark energy models with the CPL form of $w_0w_a$CDM, in which the dark energy equation of state is parameterized as $w(a)=w_0+(1-a)w_a$~\cite{ChevallierPolarski2001,Linder2003}. Although these are the models most widely discussed, especially in comparison with recent DESI observations, our same methods can be straightforwardly applied to other proposals to relieve cosmological tensions (see, e.g., \cite{2025PhRvD.112h3511L,2025NatAs...9.1879G,2025JCAP...09..053I, 2025JCAP...08..014C,2025PhRvD.111l3511S,2025PhRvD.112d3513S,2025PhRvD.112f3551G, 2025PhRvD.112f3548C, 2026PhRvD.113b3515P, 2026PhRvD.113d1304P,Wen:2026xlj}). 

Although we shall introduce our cosmological datatsets in Sec.~\ref{sec:data}, for concreteness of the methodological discussion here, $\{\mathrm{BAO},\mathrm{PP},\mathrm{U3},\mathrm{DES}\}$ refer to DESI DR2 BAO data, alongside Pantheon+, Union 3, and DES-Dovekie supernova surveys respectively.  

\subsection{Systematic uncertainty}
\label{sec:method/sys}
Looking at the history of measurements in physics and astronomy over the past century, it is reasonable to explore the possibility that any given measurement may have missed a potential source of uncertainty, which we may refer to as an unknown systematic error. A conservative approach to control for such systematics is to study statistical consistency of independent measurements, against model predictions. Here, we introduce a simple method to test for such systematics by rescaling the reported covariance of the measurements in each dataset. We multiply the published covariance matrix by a free parameter and check how much the cosmological constraints change, while remaining agnostic to the source (or existence) of that extra covariance. Let us begin with the standard Gaussian likelihood,
\begin{align}
    \mathcal{L}(x_i;\mu,\sigma_i^2)
    =
    \prod_{i=1}^{N}
    \frac{1}{\sqrt{2\pi\sigma_i^2}}
    \exp\left[
    -\frac{(x_i-\mu)^2}{2\sigma_i^2}
    \right],
    \label{eq:gaussianl}
\end{align}
where $\mu$ and $\sigma_i^2$ denote the mean and variance associated with measurement $i$, respectively.

For a dataset containing correlated measurements, we define the residual vector
\begin{equation}
    \Delta\mathbf{O}
    \equiv
    \mathbf{O}_{\rm obs}
    -
    \mathbf{O}_{\rm th},
\end{equation}
and denote its covariance matrix by $\mathbf{C}$. The corresponding multivariate Gaussian likelihood is
\begin{equation}
    \mathcal{L}
    =
    \frac{1}
    {\sqrt{(2\pi)^N\det\mathbf{C}}}
    \exp\left[
    -\frac{1}{2}
    \Delta\mathbf{O}^{\rm T}
    \mathbf{C}^{-1}
    \Delta\mathbf{O}
    \right],
    \label{eq:multivariate_gaussian}
\end{equation}
where $N$ denotes the number of data points.

Rather than modeling individual sources of systematic uncertainty, we parametrize their net effect directly at the likelihood level. For each dataset $X\in\{\mathrm{BAO},\mathrm{PP},\mathrm{U3},\mathrm{DES}\}$, we introduce a covariance-rescaling parameter, denoted by $s_{\rm BAO}$, $s_{\rm PP}$, $s_{\rm U3}$, or $s_{\rm DES}$, and define
\begin{equation}
    \mathbf{C}'_X
    =
    s_X\mathbf{C}_X,
    \qquad
    X\in
    \{\mathrm{BAO},\mathrm{PP},\mathrm{U3},\mathrm{DES}\}.
    \label{eq:covariance_rescaling}
\end{equation}
The parameter $s_X$ multiplies the covariance matrix rather than the standard deviation. Consequently, the corresponding standard deviations are rescaled by $\sqrt{s_X}$. For example, a covariance-rescaling value $s_X=1.7$ corresponds to an increase in the nominal standard deviations by a factor $\sqrt{1.7}\simeq1.30$, rather than by a factor of $1.7$. Because the adopted priors satisfy $s_X\geq1$, the analysis tests only for additional covariance and does not permit a reduction of the published covariance.

Covariance rescaling is related to Bayesian hyperparameter methods for weighting datasets \cite{Lahav2000Hyperparameters,Hobson2002Hyperparameters}. Unlike frameworks that also allow systematic shifts \cite{Bernal2018Conservative}, we consider covariance inflation and effective cross-survey correlations.

The rescaled covariance modifies both the quadratic and normalization terms of the likelihood. Omitting only additive constants that are independent of the sampled parameters, we obtain
\begin{equation}
    -2\ln\mathcal{L}_X
    =
    \Delta\mathbf{O}_X^{\rm T}
    \left(\mathbf{C}'_X\right)^{-1}
    \Delta\mathbf{O}_X
    +
    \ln\det\mathbf{C}'_X
    +
    \mathrm{const.}
    \label{eq:systematic_likelihood}
\end{equation}
The log-determinant term provides the normalization penalty associated with increasing the covariance and prevents an arbitrarily large value of $s_X$ from being preferred solely because it reduces the quadratic term.

We treat each $s_X$ as a nuisance parameter and vary it simultaneously with the cosmological parameters. We adopt the uniform priors
\begin{equation}
    s_{\rm PP},\,
    s_{\rm U3},\,
    s_{\rm DES}
    \sim
    \mathcal{U}(1,5),
\end{equation}
and
\begin{equation}
    s_{\rm BAO}
    \sim
    \mathcal{U}(1,9).
\end{equation}
The baseline value $s_X=1$ recovers the published covariance matrix, whereas $s_X>1$ allows for additional covariance beyond the reported uncertainties. Because the adopted priors are bounded below at unity, this analysis tests only for additional covariance and does not allow for a reduction of the published uncertainties.

This approach provides a simple and agnostic diagnostic of whether 
residual systematic uncertainty can affect the inferred cosmological constraints, without requiring a detailed model for its physical origin.

\subsection{Correlations}
\label{sec:method/cor}

Motivated by the need to combine heterogeneous datasets consistently, we extend the covariance-rescaling framework introduced in Sec.~\ref{sec:method/sys} to include possible correlations between different surveys. We consider three increasingly general covariance prescriptions, which differ in their treatment of survey-specific uncertainty rescaling and inter-survey correlations.

For the three supernova compilations considered here, we describe the inter-survey correlation structure using three pairwise coefficients: \(r_{\rm PP,U3}\), \(r_{\rm PP,DES}\), and \(r_{\rm U3,DES}\). These parameters control the effective cross-correlations between Pantheon+, Union3, and DES--Dovekie, respectively.

To construct the cross-covariance blocks, we define a covariance factor \(\mathbf{L}_X\) for each dataset \(X\) through
\begin{equation}
\mathbf{L}_X\mathbf{L}_X^{\rm T}=s_X\mathbf{C}_X,\qquad X\in\{\mathrm{PP},\mathrm{U3},\mathrm{DES}\},
\label{eq:covariance_square_root}
\end{equation}
where \(\mathbf{C}_X\) denotes the reported covariance matrix and \(s_X\) is the covariance-rescaling parameter introduced in Sec.~\ref{sec:method/sys}. When survey-dependent covariance rescaling is not included, we set \(s_X=1\). This construction ensures that the cross-covariance amplitudes scale consistently with the uncertainties in the corresponding diagonal covariance blocks.

After the survey-specific data vectors have been projected onto the common redshift basis described in Sec.~\ref{sec:sn/method} and Appendix~\ref{app:binning}, we construct the cross-covariance blocks from their covariance factors. Because DES--Dovekie contributes 19 populated entries on the common 22-index basis, we denote by \(P_{\rm DES}\in\mathbb{R}^{22\times19}\) the corresponding embedding matrix. In the expressions below, \(C_{\rm DES}\) and \(L_{\rm DES}\) refer to the active \(19\times19\) DES--Dovekie covariance block. The three cross-covariance blocks are constructed as
\begin{equation}
\begin{aligned}
\mathbf{C}_{\rm PP,U3} &= r_{\rm PP,U3}\mathbf{L}_{\rm PP}\mathbf{L}_{\rm U3}^{\rm T},\\
\mathbf{C}_{\rm PP,DES} &= r_{\rm PP,DES}\mathbf{L}_{\rm PP}\mathbf{P}_{\rm DES}\mathbf{L}_{\rm DES}^{\rm T},\\
\mathbf{C}_{\rm U3,DES} &= r_{\rm U3,DES}\mathbf{L}_{\rm U3}\mathbf{P}_{\rm DES}\mathbf{L}_{\rm DES}^{\rm T}.
\end{aligned}
\label{eq:sn_crosscovariances}
\end{equation}
This prescription controls the overall amplitude of each cross-covariance block without assuming a specific physical origin for the inter-survey correlation.

The resulting joint supernova covariance matrix is
\begin{equation}
\mathbf{C}_{\rm SN}=
\begin{pmatrix}
s_{\rm PP}\mathbf{C}_{\rm PP} & \mathbf{C}_{\rm PP,U3} & \mathbf{C}_{\rm PP,DES}\\
\mathbf{C}_{\rm PP,U3}^{\rm T} & s_{\rm U3}\mathbf{C}_{\rm U3} & \mathbf{C}_{\rm U3,DES}\\
\mathbf{C}_{\rm PP,DES}^{\rm T} & \mathbf{C}_{\rm U3,DES}^{\rm T} & s_{\rm DES}\mathbf{C}_{\rm DES}
\end{pmatrix}.
\label{eq:joint_sn_covariance}
\end{equation}

Within this common covariance framework, we consider three nested models. In the single-correlation model, the published within-survey covariances are retained by setting \(s_X=1\), and a common correlation coefficient is imposed such that \(r_{\rm PP,U3}=r_{\rm PP,DES}=r_{\rm U3,DES}=r\). In the three-correlation model, we retain \(s_X=1\) but allow \(r_{\rm PP,U3}\), \(r_{\rm PP,DES}\), and \(r_{\rm U3,DES}\) to vary separately. Finally, in the full model adopted in our main analysis, the three pairwise SN correlation coefficients and all four covariance-rescaling parameters, $s_{\rm PP}$, $s_{\rm U3}$, $s_{\rm DES}$, and $s_{\rm BAO}$, are varied simultaneously. The first three parameters rescale the corresponding SN covariance blocks, whereas $s_{\rm BAO}$ independently rescales the DESI DR2 BAO covariance.

Each pairwise coefficient is assigned a uniform prior over $[-1,1]$. For compactness, we define $r_{12}\equiv r_{\mathrm{PP,U3}}$, $r_{13}\equiv r_{\mathrm{PP,DES}}$, and $r_{23}\equiv r_{\mathrm{U3,DES}}$. The three coefficients are additionally required to define a positive-definite correlation matrix,
\begin{equation}
R=
\begin{pmatrix}
1 & r_{\mathrm{PP,U3}} & r_{\mathrm{PP,DES}}\\
r_{\mathrm{PP,U3}} & 1 & r_{\mathrm{U3,DES}}\\
r_{\mathrm{PP,DES}} & r_{\mathrm{U3,DES}} & 1
\end{pmatrix}.
\label{eq:correlation-matrix}
\end{equation}
The effective joint prior is consequently uniform within the positive-definite region of the original three-dimensional prior cube. The sampled coefficients should be interpreted as phenomenological, basis-dependent effective correlations rather than direct physical Pearson correlations between individual supernovae. We complement the marginalized posterior constraints with numerical minimization of the effective $\chi^2$, retaining the native likelihood terms, the Planck calibration contribution, and the parameter-dependent covariance log-determinants. For each covariance prescription, we quantify the relative best-fit improvement by subtracting the best-found effective $\chi^2$ of $w_0w_a\mathrm{CDM}$ from that of $\Lambda\mathrm{CDM}$. Within each cosmological model, we also subtract the free-scale $\chi^2$  from the fixed-scale $\chi^2$ to quantify the improvement obtained by allowing covariance rescaling. Finally, we compare the relative best-fit improvements under the fixed- and free-scale prescriptions to assess whether covariance rescaling reduces the relative fit preference for $w_0w_a\mathrm{CDM}$. All differences are evaluated within the same data combination and likelihood-normalization convention. For comparison with the convention used in the DESI analysis~\cite{DESI2025DR2BAO}, we also quote nominal Gaussian-equivalent significances obtained by evaluating the upper-tail probability $p$ of a $\chi^2$ distribution with two degrees of freedom at the effective $\chi^2$ difference between the two cosmological models and converting it to $Z=\Phi^{-1}(1-p/2)$, where \(\Phi\) is the standard normal cumulative distribution function. This conversion assumes the regularity conditions of the two-degree-of-freedom Wilks approximation. In the free-covariance analyses, the lower boundaries on the rescaling parameters may affect this approximation; we therefore report these significances as nominal diagnostics and retain the effective $\chi^2$ differences as the primary fit comparisons.

\section{Datasets}
\label{sec:data}
\subsection{Planck-lite CMB and DESI DR2 BAO data}

\begin{table*}[t]
\centering\small\setlength{\tabcolsep}{6pt}
\renewcommand{\arraystretch}{1.12}
\begin{tabular}{@{}lll@{}}
\hline\hline Probe & Likelihood or data product & Reference \\ \hline 
CMB & Planck 2018 low-$\ell$ TT/EE + high-$\ell$ Plik TTTEEE-lite & \cite{Planck2018Likelihoods,Planck2018Parameters} \\ BAO & DESI DR2 BAO & \cite{DESI2025DR2BAO} \\ SN & Pantheon+ & \cite{PantheonPlusData,PantheonPlusCosmology} \\ SN & Union3 & \cite{Union3} \\ SN & DES--Dovekie & \cite{DovekieCalibration,DESDovekie} \\ \hline\hline\end{tabular}
\caption{Cosmological likelihoods and supernova data products used in this analysis.}
\label{tab:data_summary}
\end{table*}

 We use the Planck-lite CMB likelihood, consisting of the Planck 2018 low-$\ell$ TT and EE likelihoods together with the high-$\ell$ Plik TTTEEE-lite likelihood, as implemented in Cobaya \cite{Cobaya}. The high-$\ell$ Plik-lite likelihood is pre-marginalized over the corresponding foreground, calibration, and beam nuisance parameters, substantially reducing the dimensionality of the sampled parameter space. We do not include the Planck CMB lensing likelihood. Throughout this work, ``Planck-lite'' refers specifically to this low-$\ell$ TT and EE plus high-$\ell$ Plik TTTEEE-lite combination and not to the full Planck baseline likelihood.

The CMB likelihood provides strong constraints on early-Universe cosmology and serves as an anchor for late-time probes such as BAO and supernovae \cite{Planck2018Parameters}. In our analysis, it is used to break degeneracies between cosmological parameters and to provide a consistent reference for assessing potential tensions between datasets. We vary the standard cosmological parameters jointly with the nuisance parameters introduced in our analysis.

In addition to Planck data, we include baryon acoustic oscillation (BAO) measurements from the Dark Energy Spectroscopic Instrument (DESI) Data Release 2~\cite{DESI2025DR2BAO}. This dataset provides distance measurements inferred from large-scale structure observations over a wide redshift range.

In particular, the BAO signal is used to extract constraints on distance indicators such as the comoving angular diameter distance $D_M(z)$, the Hubble parameter $H(z)$, and/or the volume-averaged distance $D_V(z)$, typically normalized by the sound horizon at the drag epoch, $r_d$ \cite{DESI2025DR2BAO}. These measurements are provided at multiple effective redshifts, along with their associated covariance matrices.

The DESI DR2 likelihood is constructed assuming Gaussian-distributed uncertainties, with the covariance matrix encoding both statistical errors and survey-specific systematic effects as reported by the collaboration. In our analysis, we use the publicly available measurements and covariance matrices, and incorporate them within the Gaussian likelihood framework described in Sec.~\ref{sec:method}.

We also allow the DESI covariance matrix to be rescaled by the parameter $s_{\rm BAO}$. This allows us to test whether extra BAO covariance changes the cosmological results. This enables us to assess the impact of potential residual systematics on the inferred cosmological constraints and their consistency with other probes.

\subsection{Supernova data}
 We use three Type Ia supernova compilations: Pantheon+~\cite{PantheonPlusData,PantheonPlusCosmology}, Union3~\cite{Union3}, and DES--Dovekie~\cite{DovekieCalibration,DESDovekie}. These compilations differ in survey composition, photometric calibration, sample selection, light-curve modeling, bias corrections, and treatments of intrinsic scatter. Pantheon+ combines observations from multiple low- and high-redshift surveys \cite{PantheonPlusData,PantheonPlusCosmology}, Union3 analyzes a heterogeneous compilation through the UNITY1.5 Bayesian framework \cite{Union3}, and DES--Dovekie reanalyzes the DES five-year and low-redshift samples using an updated calibration and light-curve model \cite{DovekieCalibration,DESDovekie}.

Potential residual differences may arise from photometric cross-calibration, light-curve-model training, selection effects, redshift and peculiar-velocity corrections, host-galaxy correlations, and shared low-redshift observations or calibration inputs. Published analyses do not yield identical \(w_0\)–\(w_a\) constraints or identical preferences relative to \(\Lambda\)CDM. Union3 reports a weak \(1.7\)–\(2.6\sigma\) tension with \(\Lambda\)CDM, while the updated DES--Dovekie calibration reduces the reported preference for evolving dark energy relative to the original DES-SN5YR result \cite{DES5YR,DovekieCalibration,DESDovekie}. These differences motivate a joint robustness test, but do not by themselves demonstrate the presence of unmodeled systematics.

For the joint likelihood, the native 22-point Union3 redshift representation defines the common basis \cite{Union3}. Pantheon+ is restricted to \(z_{\rm HD}>0.01\), retaining 1590 entries \cite{PantheonPlusData}, and is projected onto this basis using the linear operator described in Appendix~\ref{app:binning} \cite{DovekieCalibration,DESDovekie}. DES--Dovekie is represented using the same nominal indexing, but only 19 populated bins enter its residual vector and covariance matrix. The three unpopulated high-redshift entries are retained only for index alignment and are never treated as measurements. The resulting joint residual vector therefore contains \(22_{\rm PP}+22_{\rm U3}+19_{\rm DES}=63\) observed entries.

\section{BAO treatment}
\label{sec:bao}
\begin{figure*}
    \centering
    \includegraphics[width=0.7\textwidth]{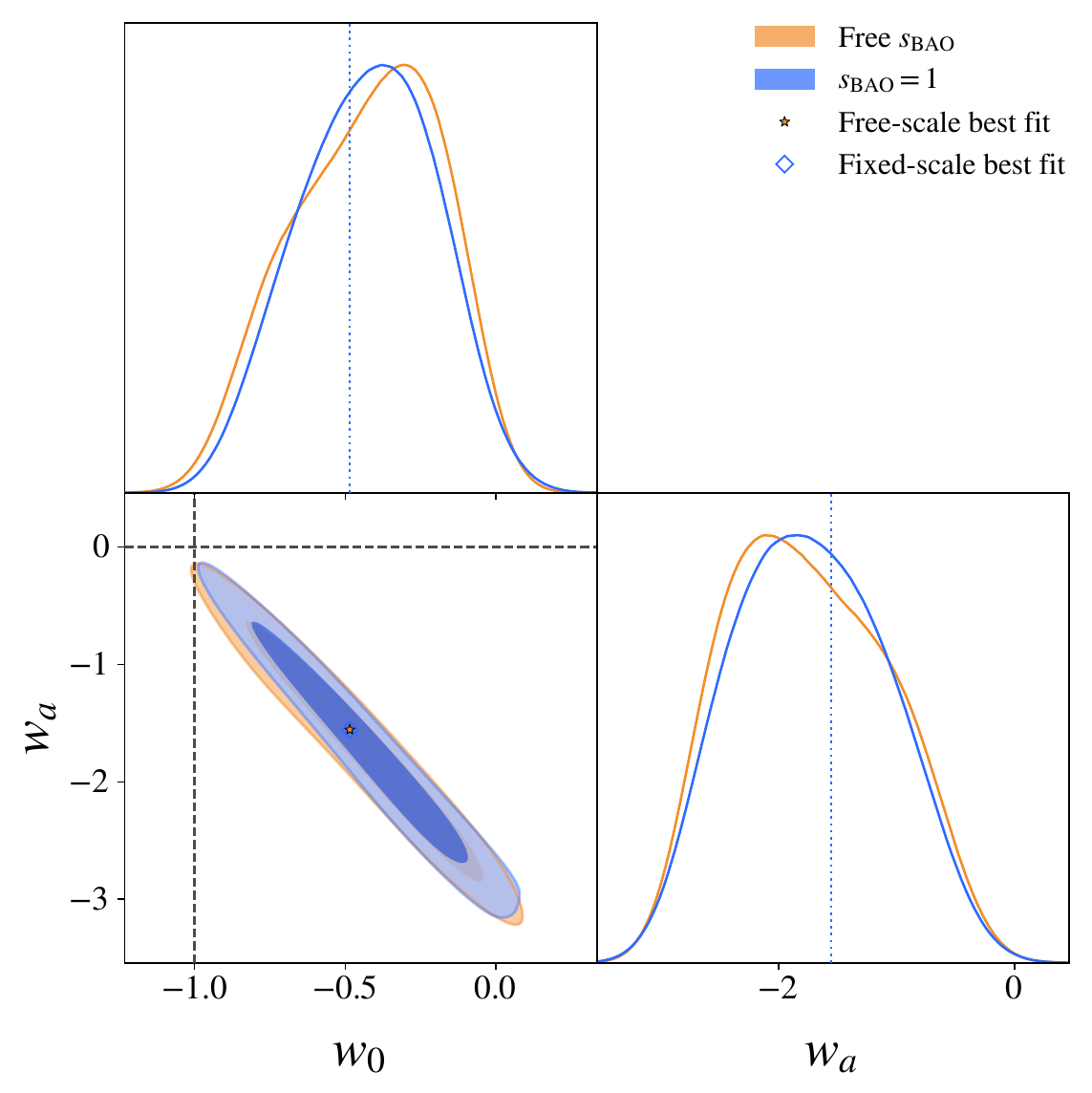}
    \caption{Constraints on $(w_0,w_a)$ under $w_0w_a$CDM from DESI DR2 BAO combined with Planck-lite. Orange contours show the analysis with $s_{\rm BAO}$ varied, while blue contours show the baseline analysis with $s_{\rm BAO}=1$. The inner and outer contours enclose the 68\% and 95\% credible regions, respectively. The dashed lines mark the $\Lambda$CDM point $(w_0,w_a)=(-1,0)$. Allowing $s_{\rm BAO}$ to vary produces a mild broadening along the principal degeneracy direction. The markers indicate the best-found effective-$\chi^2$ minima of the corresponding fixed- and free-\(s_{\rm BAO}\) analyses; both lie within their respective 68\% credible contours.}
    \label{fig:baosys}
\end{figure*}

We first apply this method to DESI DR2 BAO combined with the Planck-lite CMB likelihood. Fig.~\ref{fig:baosys} compares the baseline analysis with $s_{\rm BAO}=1$ to the extended analysis in which $s_{\rm BAO}$ is allowed to vary. Table~\ref{tab:baosys} reports posterior medians and central 68\% credible intervals for the cosmological parameters, and the boundary-corrected marginalized mode and 95\% upper credible limit for $s_{\rm BAO}$.

Allowing $s_{\rm BAO}$ to vary produces only a mild broadening of the cosmological contours. The marginalized posterior is broad and extends to the lower prior boundary at $s_{\rm BAO}=1$. Because the adopted prior is one-sided, $s_{\rm BAO}\sim\mathcal{U}(1,9)$, this result quantifies the amount of additional BAO covariance compatible with the data; it does not establish either that the published covariance is underestimated or that no rescaling is required.

\begin{table}[t]
\centering\footnotesize\setlength{\tabcolsep}{5pt}\renewcommand{\arraystretch}{1.15}
\begin{tabular}{@{}lcc@{}}
\hline\hline 
Parameter & $s_{\rm BAO}=1$ & Free $s_{\rm BAO}$ \tabularnewline \hline $w_0$ & $-0.43^{+0.24}_{-0.21}$ & $-0.48^{+0.27}_{-0.21}$ \tabularnewline $w_a$ & $-1.70\pm0.62$ & $-1.59^{+0.64}_{-0.76}$ \tabularnewline $\Omega_m$ & $0.352\pm0.022$ & $0.348^{+0.026}_{-0.024}$ \tabularnewline $s_{\rm BAO}$ & $1\ \mathrm{(fixed)}$ & $1.00\;(<2.69)$ 
\tabularnewline 
\hline\hline\end{tabular}
\caption{Marginalized posterior medians and 68\% central credible intervals under $w_0w_a$CDM for DESI DR2 BAO combined with Planck-lite. The two columns compare the baseline analysis with $s_{\rm BAO}=1$ and the extended analysis with $s_{\rm BAO}$ free. Allowing $s_{\rm BAO}$ to vary produces only mild changes in the marginalized cosmological constraints. Cosmological parameters are posterior medians with central 68\% credible intervals; $s_{\rm BAO}$ is the boundary-corrected marginalized mode with its 95\% upper credible limit in parentheses}
\label{tab:baosys}
\end{table}

In the individual-SN diagnostic analyses presented below, we fix $s_{\rm BAO}=1$ in order to isolate the effect of each SN covariance-rescaling parameter. In the final DESI DR2 BAO + Planck-lite + joint-SN analysis, however, $s_{\rm BAO}$ is varied simultaneously with the SN correlation and covariance-rescaling parameters.

The best-fit comparison quantifies the effect of BAO covariance rescaling. For DESI DR2 BAO combined with Planck-lite, extending $\Lambda\mathrm{CDM}$ to $w_0w_a\mathrm{CDM}$ lowers the best-found effective $\chi^2$ by $9.25$ when $s_{\mathrm{BAO}}=1$ and by $9.25$ when $s_{\mathrm{BAO}}$ is varied. Using a nominal $\chi^2$ distribution with two degrees of freedom and expressing the resulting tail probability as a two-sided Gaussian-equivalent significance, both correspond to approximately $2.58\sigma$. Thus, allowing BAO covariance rescaling does not improve the relative best-fit, leaving the relative fit preference for $w_0w_a\mathrm{CDM}$ essentially unchanged. The corresponding best-fit points are shown in Fig.~\ref{fig:baosys}, and the fit comparisons are summarized in Table~\ref{tab:bestfit_comparison}.

\section{SN treatment}
\label{sec:sn}
The BAO covariance-rescaling test only slightly changes the cosmological constraints. We now apply the same method to the Type Ia supernova datasets. Unlike BAO surveys, modern SN compilations are constructed from heterogeneous observational samples with differing calibration procedures, selection criteria, and light-curve fitting methodologies. As a result, residual survey-dependent systematics and inter-survey correlations may play a more significant role in the inferred cosmological constraints.

\begin{figure*}
    \centering
    \includegraphics[width=0.7\textwidth]{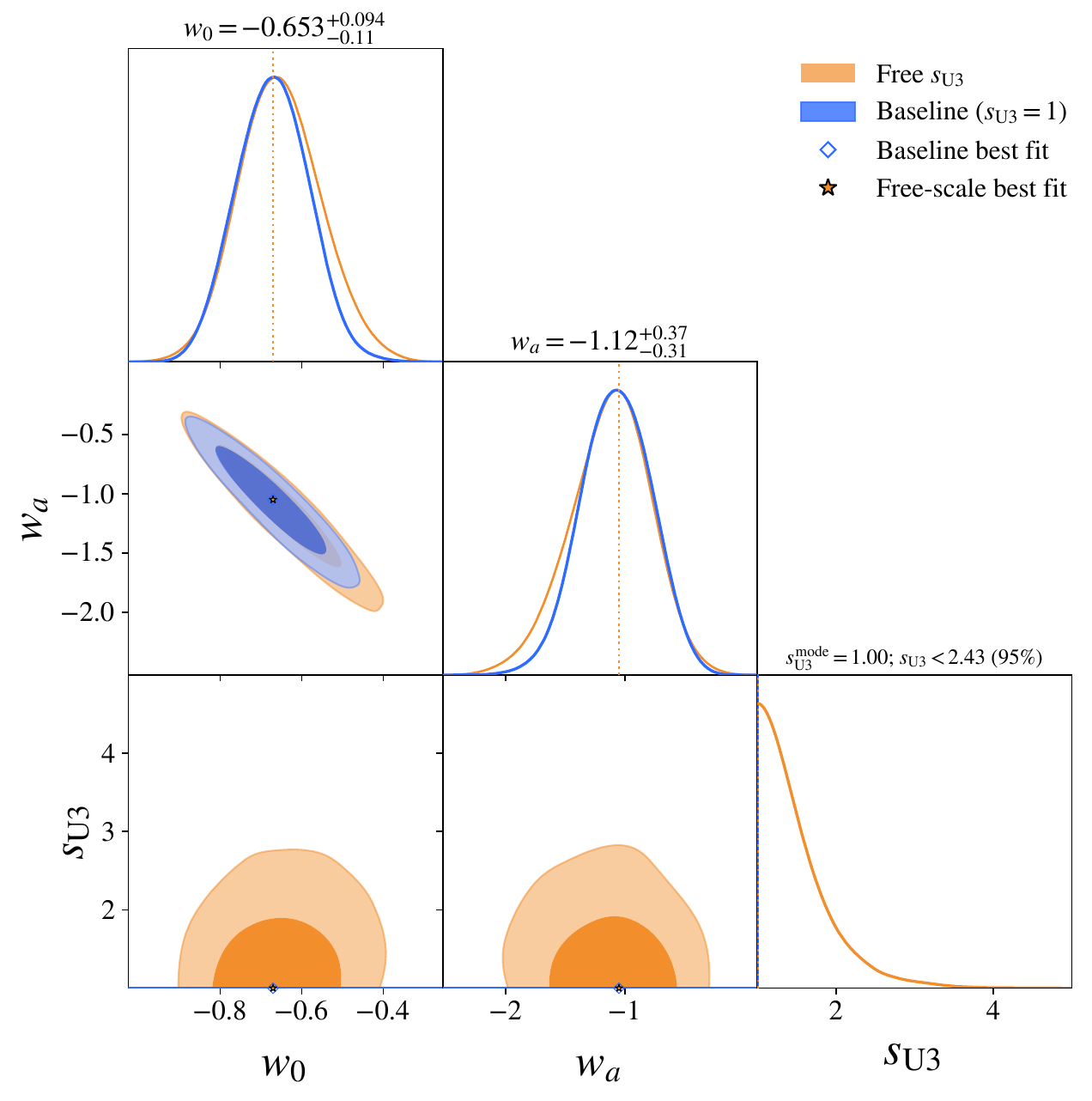}
    \caption{Posterior constraints on $w_0$, $w_a$, and $s_{\rm U3}$ under $w_0w_a$CDM from DESI DR2 BAO + Planck-lite + Union3. Orange contours show the analysis with free $s_{\rm U3}$, while blue contours show the baseline with $s_{\rm U3}=1$. Inner and outer contours enclose 68\% and 95\% credible regions. The baseline scale is represented by a line at $s_{\rm U3}=1$, since it is fixed rather than sampled. The scale density, its mode, and its 95\% upper credible limit use the same boundary-corrected posterior estimate. The open blue diamond and orange star mark the baseline and free-scale best-fit points, respectively. Allowing covariance rescaling mildly broadens the cosmological constraints. The dotted lines indicate the best-fit parameter values for the corresponding analyses.}
    \label{fig:union3sys}
\end{figure*}

\begin{table}[htbp]
\centering
\small
\setlength{\tabcolsep}{6pt}
\renewcommand{\arraystretch}{1.15}
\begin{tabular}{lcc}
\hline\hline
Parameter & Baseline & Free $s_{\mathrm{U3}}$ \\
\hline
$w_0$ & $-0.67^{+0.09}_{-0.09}$ & $-0.653^{+0.009}_{-0.12}$ \\
$w_a$ & $-1.07^{+0.30}_{-0.30}$ & $-1.12^{+0.37}_{-0.31}$ \\
$\Omega_m$ & $0.327^{+0.009}_{-0.009}$ & $0.329^{+0.010}_{-0.010}$ \\
$s_{\rm U3}$ & $1$ (fixed) & $1.00\;(<2.43)$ \\
\hline\hline
\end{tabular}
\caption{Constraints in $w_0w_a$CDM from Union3 combined with Planck-lite and DESI DR2 BAO. Cosmological parameters are summarized by posterior medians and central 68\% credible intervals. For $s_{\rm U3}$, we report the boundary-corrected one-dimensional marginalized posterior mode, rounded to two decimal places, with the 95\% upper credible limit in parentheses. The baseline fixes $s_{\rm U3}=1$, while the extended analysis varies $s_{\rm U3}$ with $C'_{\rm U3}=s_{\rm U3}C_{\rm U3}$. Both analyses fix $s_{\rm BAO}=1$.}
\label{tab:union3sys_cosmo}
\end{table}

The individual-survey covariance-rescaling constraints are summarized together with the joint SN (without CMB or BAO) and full-combination results in Table~\ref{tab:sn_nuisance_summary}.
  
\begin{figure*}
    \centering
    \includegraphics[width=\textwidth]{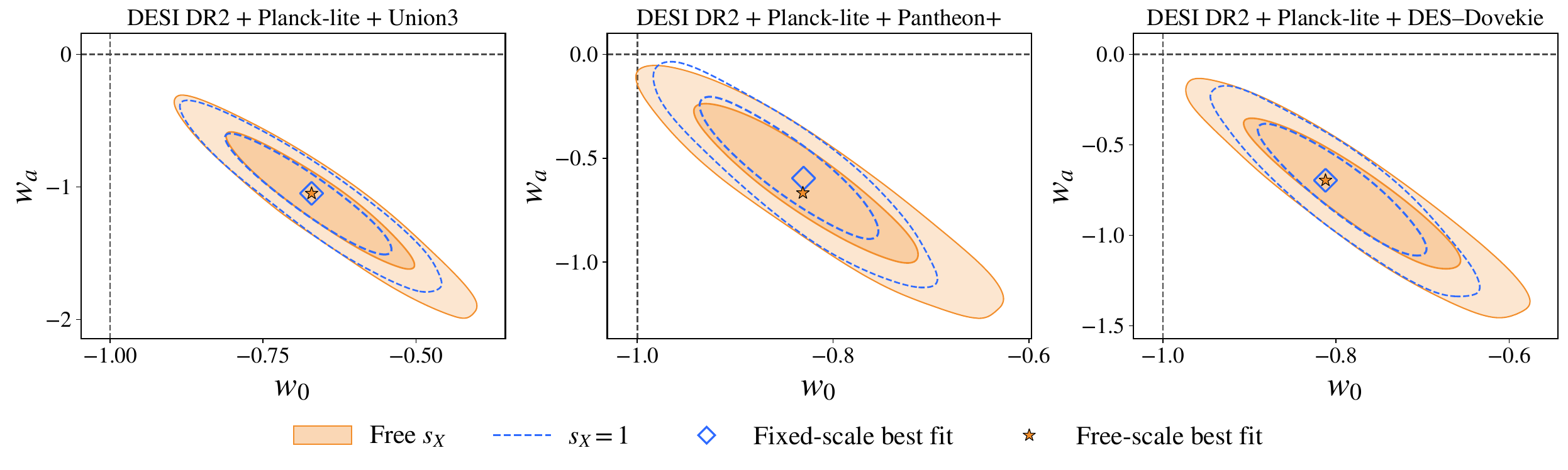}
    \caption{Constraints on $(w_0,w_a)$ under $w_0w_a$CDM from DESI DR2 BAO and Planck-lite, combined separately with Union3 (left), Pantheon+ (center), and DES--Dovekie (right). Orange filled contours show the analyses with free $s_X$, while blue dashed contours show the baseline analyses with $s_X=1$. The contours show the 68\% and 95\% credible regions, and the dotted lines mark the $\Lambda$CDM point $(w_0,w_a)=(-1,0)$. Allowing $s_X$ to vary slightly broadens the contours but does not clearly change their directions or preferred regions. The markers indicate the best-found effective-$\chi^2$ minima for the fixed- and free-\(s_X\) analyses. In all three panels, the best-fit points lie within the corresponding 68\% credible contours.}
    \label{fig:wwa_sys_sn}
\end{figure*}

We now apply the framework developed in Sec.~\ref{sec:method} to supernova datasets. As a first example, we next test the effect of covariance rescaling for Union3 combined with DESI DR2 BAO and Planck-lite. We can compare the baseline analysis with $s_{\rm U3}=1$ to the analysis in which $s_{\rm U3}$ is allowed to vary. The cosmological constraints are shown in Fig.~\ref{fig:union3sys}, and the corresponding parameter constraints are listed in Table~\ref{tab:union3sys_cosmo}. The same comparison for $(w_0,w_a)$ constraints for other SN surveys is shown in Fig. (\ref{fig:wwa_sys_sn}).

The boundary-corrected marginalized posterior for $s_{\rm U3}$ peaks close to the lower prior boundary, with mode $s_{\rm U3}\simeq1.004$ and 95\% upper credible limit $s_{\rm U3}<2.43$. Marginalizing over its broad tail toward larger scales can broaden the cosmological contours even though the mode is close to unity. Relative to the baseline analysis, the posterior in the \((w_0,w_a)\) plane extends further toward larger \(w_0\) and more negative \(w_a\), following the existing degeneracy direction, while the central regions remain substantially overlapping. As listed in Table~\ref{tab:union3sys_cosmo}, the posterior medians change from \(w_0=-0.67\) to \(-0.65\), from \(w_a=-1.07\) to \(-1.10\), and from \(\Omega_m=0.327\) to \(0.329\). The best-found effective $\chi^2$ difference, $\chi^2_{\Lambda\mathrm{CDM}}-\chi^2_{w_0w_a\mathrm{CDM}}$, decreases from $15.00$ with fixed scales to $14.27$ with free scales, corresponding to nominal Gaussian-equivalent significances of \(3.45\sigma\) and \(3.35\sigma\), respectively. This represents a modest reduction in the relative best-fit advantage of \(w_0w_a\mathrm{CDM}\). Since the prior starts at \(s_{\mathrm{U3}}=1\), the rescaling posterior should not by itself be interpreted as evidence that the published Union3 covariance is underestimated.

The extra covariance allowed by the data may be related to several known challenges in supernova analyses, including calibration uncertainties, selection effects, differences between surveys, and differences in light-curve standardization methods \cite{DovekieCalibration,PantheonPlusCosmology,Union3}. Since Union3 combines data from several observational programs with different analysis pipelines, some differences may remain after the standard corrections are applied.

The individual-survey analyses give boundary-corrected marginalized posterior modes $s_{\rm U3}=1.00$, $s_{\rm PP}=1.30$, and $s_{\rm DES}=1.00$, with respective 95\% upper credible limits of $2.43$, $2.99$, and $2.84$. These broad posteriors retain substantial support near $s_X=1$ and quantify the additional covariance allowed within the adopted one-sided rescaling model; they do not establish that the published uncertainties are too small. We next consider the joint-SN likelihood with the inter-survey correlation framework introduced in Sec.~\ref{sec:method}.

The impact of the additional systematic freedom on the $(w_0, w_a)$ constraints is illustrated in Fig.~\ref{fig:wwa_sys_sn} for each of the three compilations separately. Allowing $s_X$ to vary makes the contours slightly wider for all three SN compilations. However, their overall shapes, directions, and preferred regions remain almost unchanged. For these data combinations, the visible effect of marginalizing over the covariance-rescaling parameter is primarily a broadening of the cosmological posterior.

\subsection{Combination of SN data}
\label{sec:sn/method}
We now apply the same covariance-rescaling method to the joint analysis of the three SN compilations.

\subsubsection{Binning strategy}

To construct the phenomenological cross-covariance blocks, we represent Pantheon+, Union3, and DES--Dovekie on a common index-aligned basis defined by the native 22-point Union3 redshift grid. The common basis provides a consistent correspondence between survey bins but does not assume that measurements assigned to the same bin are identical. Pantheon+ is compressed from the 1590 supernovae satisfying $z_{\mathrm{HD}}>0.01$ onto all 22 bins, Union3 is retained in its native representation, and only the 19 populated DES--Dovekie bins enter the likelihood. The active joint supernova residual therefore contains $22_{\mathrm{PP}}+22_{\mathrm{U3}}+19_{\mathrm{DES}}=63$ observed entries.

For each compilation $X$, the data vector, covariance, and raw theoretical prediction are projected consistently according to
\begin{equation}
\begin{aligned}
\widetilde{\boldsymbol{d}}_X &= W_X\boldsymbol{d}_X, &
\widetilde C_X &= W_XC_XW_X^{\mathsf T},\\
\widetilde{\boldsymbol{\mu}}_X(\boldsymbol{\theta}) &= W_X\boldsymbol{\mu}_X^{\mathrm{raw}}(\boldsymbol{\theta}).
\end{aligned}
\label{eq:main-sn-projection}
\end{equation}
The Pantheon+ weights are obtained from local generalized least-squares combinations within each redshift bin, while the complete selected covariance matrix is propagated through Eq.~(\ref{eq:main-sn-projection}). The theoretical prediction is evaluated at the original CMB-frame and heliocentric redshifts before projection and is not approximated by evaluating the model only at the nominal or effective bin redshifts. The three empty high-redshift DES--Dovekie bins are retained only in the stored common-grid representation and are excluded from the residual vector, covariance matrix, and inter-survey cross-covariance blocks.

The final joint analysis uses the 22-bin Pantheon+ construction because it matches the native Union3 indexing required by the cross-covariance model. Comparisons with the raw Pantheon+ likelihood show that alternative 24- and 26-bin compressions provide closer agreement, particularly along the weakly constrained directions of $w_0w_a$CDM, while leaving the inferred cosmological region and the conclusions of the analysis unchanged within the marginalized uncertainties. These alternative constructions are therefore used only to assess compression sensitivity and are not included in the final joint likelihood. The complete bin definitions, projection procedure, and validation tests are provided in Appendix~\ref{app:binning}.

\begin{figure*}[t]
\centering
\includegraphics[width=\textwidth]{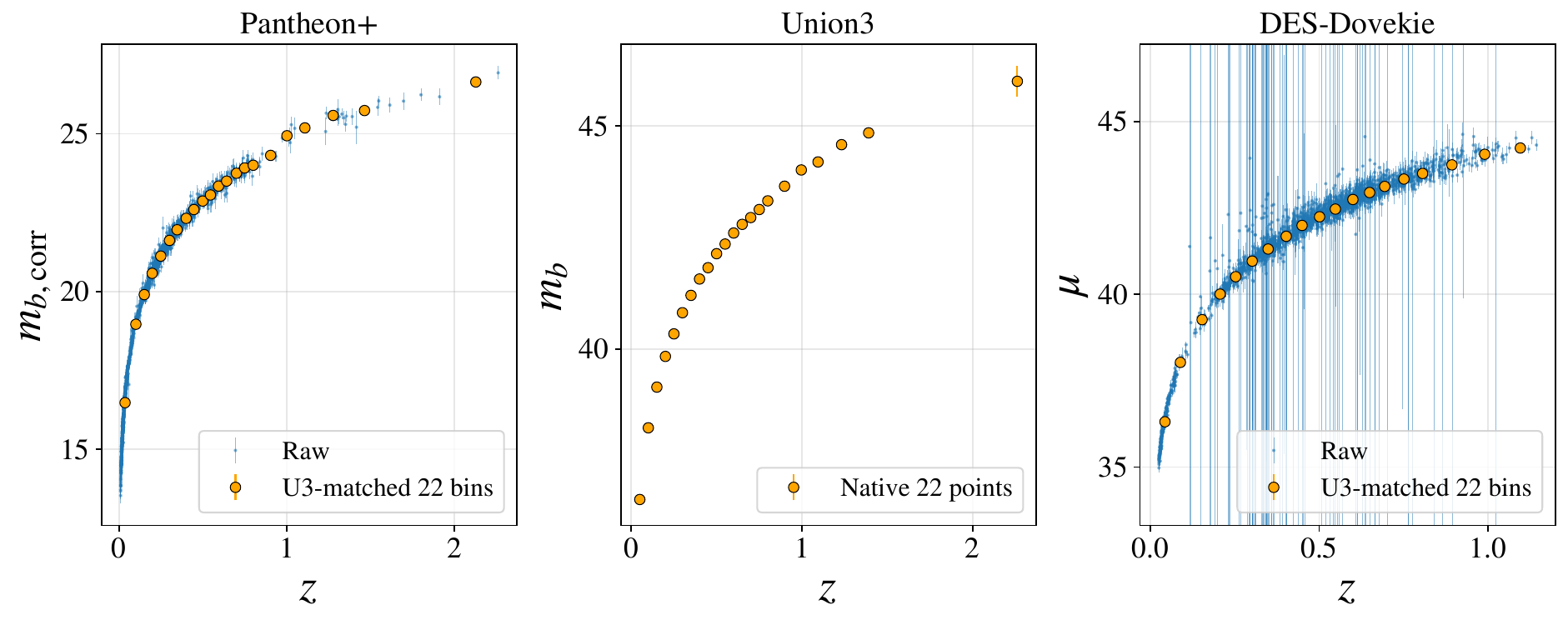}
\caption{Comparison of the native SN observables with their common-grid representations. The left panel shows the 1590 Pantheon+ measurements retained after applying $z_{\rm HD}>0.01$ together with their Union3-matched 22-bin projection. The middle panel shows Union3 in its native 22-point representation, which is retained without further compression. The right panel shows the original DES--Dovekie measurements and their Union3-matched representation. Nineteen of the nominal 22 DES--Dovekie bins are populated; the three empty high-redshift bins are omitted from the active likelihood.}
\label{fig:rebin}
\end{figure*}

Having established a common redshift basis for the three SN compilations, we now investigate whether effective inter-survey correlations influence the inferred cosmological constraints. Such correlations may arise from overlapping low-redshift samples, shared calibration procedures, or common sources of observational uncertainty. The correlation models and their cosmological implications are examined in the following subsection.

The dependence of the inferred SN nuisance parameters on the cosmological model is summarized in Table~\ref{tab:sn_nuisance_summary}.

\begin{figure*}[t]
\centering
\includegraphics[width=0.80\textwidth]{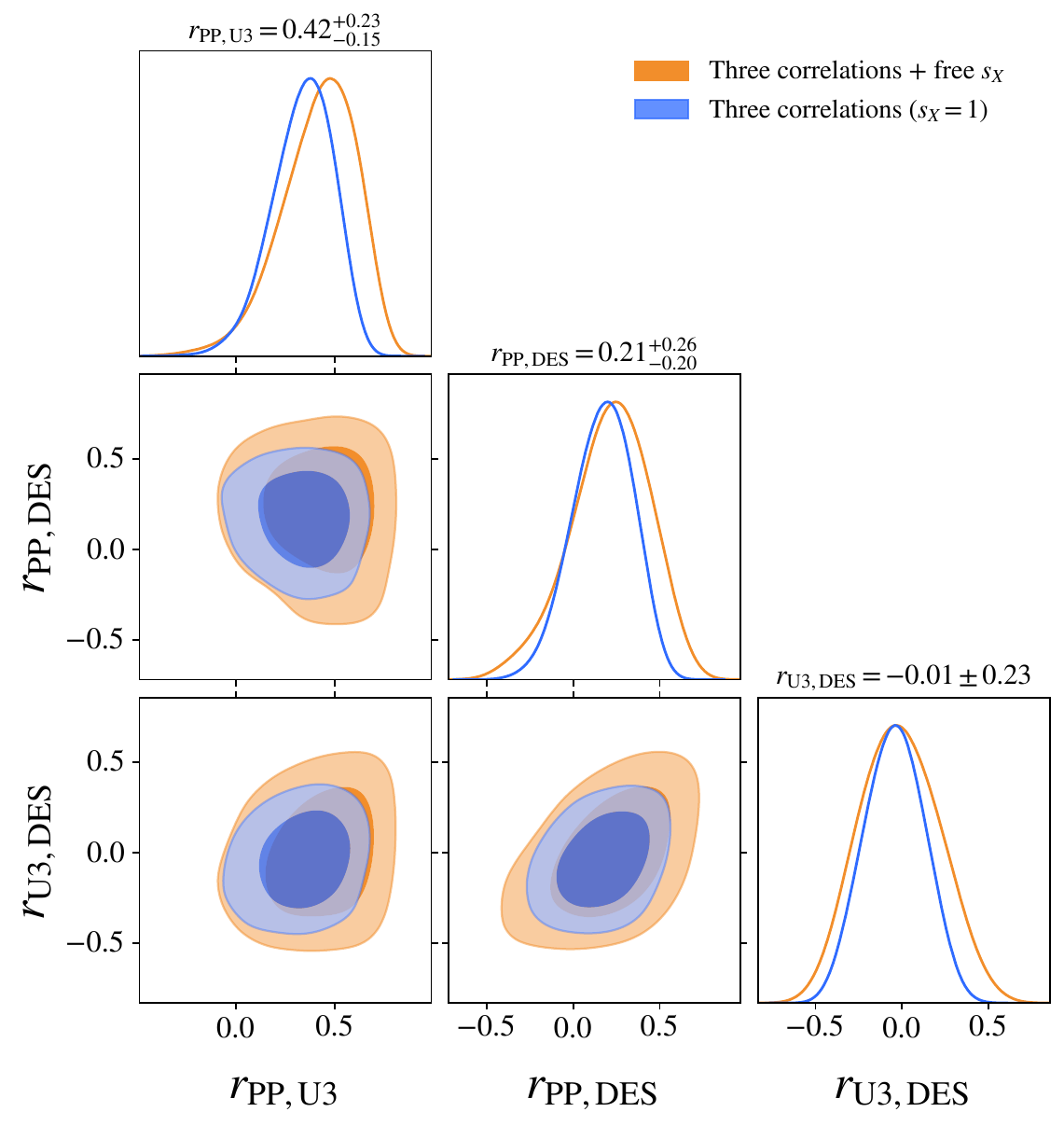}
\caption{Marginalized SN-only constraints under $\Lambda$CDM from the joint Pantheon+ + Union3 + DES--Dovekie likelihood. Blue contours show the three-correlation model with $s_X=1$, while orange contours show the $3r+3s$ model with free correlation and covariance-rescaling parameters. The contours show the 68\% and 95\% credible regions. Allowing $s_X$ to vary only slightly changes the correlation constraints.}
\label{fig:sncor_lcdm}
\end{figure*}

\begin{figure*}
    \centering
    \includegraphics[width=0.7\textwidth]{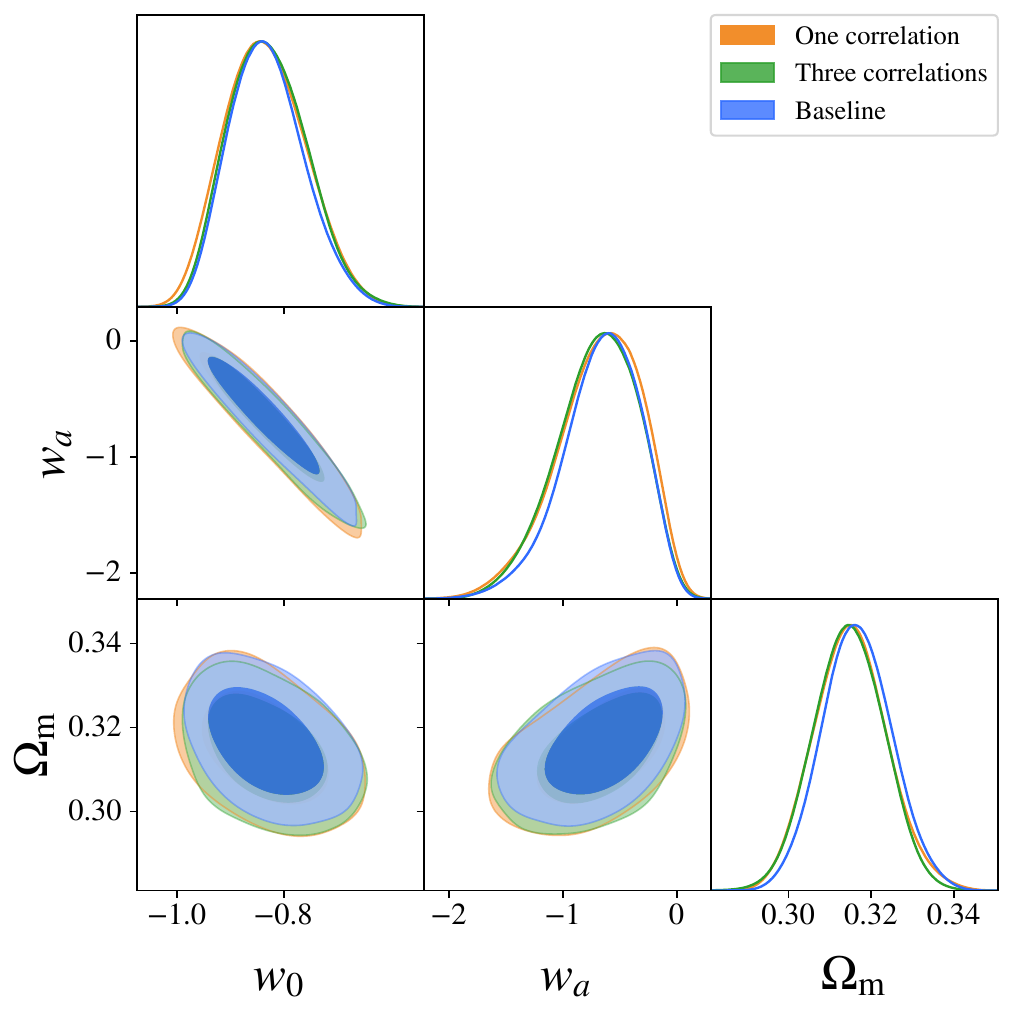}
    \caption{Cosmological constraints from the joint Pantheon+ + Union3 + DES--Dovekie likelihood combined with Planck-lite under $w_0w_a$CDM. The zero-correlation comparison fixes $r_{XY}=0$ and $s_X=1$, neglecting cross-survey covariance despite shared supernovae. The one-correlation model uses one common correlation parameter, while the three-correlation model allows the three pairwise correlations to vary separately. All covariance-rescaling parameters are fixed to one, and no DESI DR2 BAO data are included. The three sets of contours mostly overlap, showing that the correlation model has little effect on the cosmological constraints.}
    \label{fig:sncor}
\end{figure*}

\subsubsection{Correlation analysis}
\label{sec:sn/method/cor}

We investigate the influence of the inter-survey correlation parameters first within the SN-only likelihood and then after including external cosmological information. Figure~\ref{fig:sncor_lcdm} compares the correlation-only and full covariance models for the joint SN likelihood under $\Lambda$CDM. Figure~\ref{fig:sncor} shows the corresponding cosmological comparison after combining the SN likelihood with the Planck-lite CMB likelihood, while Table~\ref{tab:sn_nuisance_summary} summarizes the final nuisance-parameter constraints obtained from the full DESI DR2 BAO + Planck-lite + SN combination.

For the single-correlation SN + Planck-lite model, we obtain
\begin{equation}
r=0.08^{+0.11}_{-0.20},
\end{equation}
which is fully consistent with zero within the marginalized uncertainties. A single common coefficient therefore provides no evidence for an overall residual correlation shared by all three pairs of SN compilations.

The SN-only three-correlation analysis nevertheless permits different correlation patterns for the three survey pairs. Under $\Lambda$CDM, the full SN-only model gives $r_{\mathrm{PP,U3}}=0.45^{+0.21}_{-0.15}$, $r_{\mathrm{PP,DES}}=0.32^{+0.27}_{-0.19}$, and $r_{\mathrm{U3,DES}}=0.00\pm0.25$. Under $w_0w_a$CDM, the corresponding SN-only constraints are $r_{\mathrm{PP,U3}}=0.44^{+0.16}_{-0.23}$, $r_{\mathrm{PP,DES}}=0.20^{+0.24}_{-0.27}$, and $r_{\mathrm{U3,DES}}=-0.10^{+0.27}_{-0.26}$. The (Pantheon+)$\times$(Union3) coefficient shows the clearest positive effective correlation, plausibly reflecting in part their shared supernovae and common calibration inputs; the (Pantheon+)$\times$(DES--Dovekie) constraint is weaker and includes zero in its central 68\% interval under $w_0w_a$CDM, while the (Union3)$\times$(DES--Dovekie) coefficient remains consistent with zero.

For the final DESI DR2 BAO + Planck-lite + SN combination, Table~\ref{tab:sn_nuisance_summary} gives $r_{\mathrm{PP,U3}}=0.50^{+0.15}_{-0.19}$, $r_{\mathrm{PP,DES}}=0.32^{+0.20}_{-0.25}$, and $r_{\mathrm{U3,DES}}=0.07^{+0.24}_{-0.25}$ under $\Lambda$CDM. Under $w_0w_a$CDM, the corresponding values are $r_{\mathrm{PP,U3}}=0.43^{+0.17}_{-0.22}$, $r_{\mathrm{PP,DES}}=0.16^{+0.24}_{-0.26}$, and $r_{\mathrm{U3,DES}}=-0.12^{+0.27}_{-0.24}$. The (Pantheon+)$\times$(Union3) correlation remains positive in both cosmological models. The (Pantheon+)$\times$(DES--Dovekie) coefficient is positive under $\Lambda$CDM but becomes weaker and consistent with zero under $w_0w_a$CDM, while the (Union3)$\times$(DES--Dovekie) coefficient remains consistent with zero in both cases.

The correlation-only and full covariance models produce substantially overlapping cosmological contours. Allowing the covariance-rescaling parameters to vary introduces only mild broadening and no significant displacement of the preferred cosmological region. The additional covariance freedom therefore affects the nuisance-parameter sector more strongly than the inferred cosmological parameters.

\subsubsection{Adding systematics}
\label{sec:sn/method/sys}

We now combine the two phenomenological extensions introduced in the previous sections by simultaneously including inter-survey correlations and survey-dependent covariance-rescaling nuisance parameters in the joint supernova analysis. In this framework, each dataset is allowed to possess an independent covariance-rescaling parameter, while the correlations between different surveys are modeled through the covariance structure described in Sec.~\ref{sec:sn/method/cor}.

This extended setup provides the most general treatment considered in this work, allowing for both residual internal systematics within each compilation and possible correlations arising from overlapping supernovae, shared calibration procedures, or common observational effects. The cosmological parameters are therefore constrained simultaneously with the full set of nuisance parameters describing the covariance structure of the combined likelihood.

Having established the methodology in the previous sections, we now proceed to investigate the impact of these additional degrees of freedom on the inferred cosmological constraints. Fig.~\ref{fig:sncor} compares three correlation prescriptions using the joint Pantheon+ + Union3 + DES--Dovekie likelihood combined with Planck-lite. The zero-correlation comparison neglects cross-survey covariance and may therefore double-count information from shared supernovae. The one-correlation model uses a single common coefficient for the three survey pairs, while the three-correlation model varies the pairwise coefficients independently. All covariance-rescaling parameters are fixed to unity, and DESI DR2 BAO is not included in this comparison. The resulting cosmological contours substantially overlap, indicating that the alternative correlation prescriptions do not produce a clear displacement of the marginalized cosmological posterior.

We find that the inclusion of these additional nuisance parameters has a relatively small impact on the inferred cosmological constraints. In particular, the $(w_0,w_a)$ contours obtained in the different scenarios exhibit substantial overlap, indicating that the cosmological predictions remain stable under the extended covariance treatment. Although the inclusion of systematics slightly broadens the allowed parameter space, no significant shifts in the preferred cosmological region are observed.

The overlapping contours demonstrate limited sensitivity to the covariance prescriptions tested here, but do not establish consistency between the compilations or validate neglecting their shared supernovae.

\begin{figure*}
    \centering
    \includegraphics[width=\textwidth]{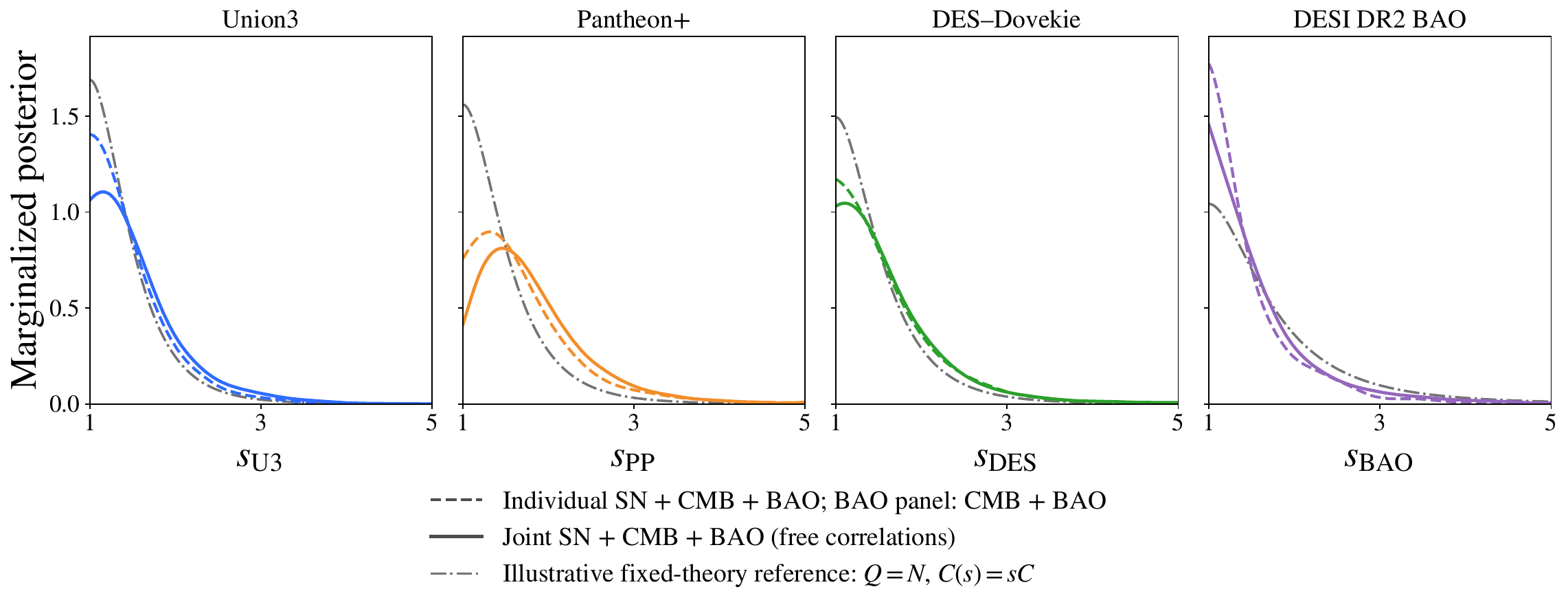}
    \caption{Boundary-corrected marginalized posteriors of the covariance-rescaling parameters under $w_0w_a$CDM. In the SN panels, dashed curves combine each SN compilation separately with Planck-lite and DESI DR2 BAO, fixing $s_{\rm BAO}=1$ and including no inter-survey correlations. Solid curves show the full Planck-lite + DESI DR2 BAO + joint-SN analysis, varying all three SN correlation coefficients, the three SN scales, and $s_{\rm BAO}$. In the BAO panel, the dashed curve uses Planck-lite + DESI DR2 BAO with free $s_{\rm BAO}$, while the solid curve uses the full combination. Grey reference curves show the Gaussian scale dependence for a fixed theoretical prediction with residual quadratic form $Q=N$, where $N$ is the number of populated SN bins in the SN panels; they are illustrative reference curves rather than simulation-calibrated null distributions.}
    \label{fig:systematics_comparison}
\end{figure*}

For comparison, Fig.~\ref{fig:systematics_comparison} includes reference curves of the form $p(s)\propto s^{-N/2}\exp[-N/(2s)]$, obtained for a fixed theoretical prediction with residual quadratic form $Q=N$. We use $N=22$ for Pantheon+ and Union3 and $N=19$ for DES--Dovekie, corresponding to the number of populated bins. In the idealized Gaussian problem with a fixed mean and a single unrestricted covariance scale, the estimator $\hat{s}=Q/N$ has standard deviation $\sqrt{2/N}$ when the true scale is unity, giving approximately $0.30$ for $N=22$ and $0.32$ for $N=19$. These values provide an indication of the fluctuations expected with a small data vector, but they do not assign a detection significance to the marginalized modes in our joint analysis, which also varies cosmological and covariance parameters and imposes $s\geq1$. The reference curves illustrate the scale dependence of the Gaussian likelihood; they are not simulation-calibrated null distributions.

The nuisance-parameter results of the full analysis are summarized in Table~\ref{tab:sn_nuisance_summary}. The $s_X$ posteriors are broad and extend above one. The (Pantheon+)$\times$(Union3) coefficient shows the clearest positive value, the (Pantheon+)$\times$(DES--Dovekie) coefficient is weaker, and the (Union3)$\times$(DES--Dovekie) coefficient remains consistent with zero.

Figure~\ref{fig:systematics_comparison} compares the covariance-rescaling posteriors obtained from the individual-SN analyses with those from the full joint analysis. For the three SN parameters, the dashed curves correspond to analyses in which each SN compilation is combined separately with DESI DR2 BAO and Planck-lite, with $s_{\rm BAO}=1$ and no inter-survey correlations. The solid curves correspond to the full joint analysis, in which the three SN correlation coefficients, the three SN covariance-rescaling parameters, and $s_{\rm BAO}$ are varied simultaneously. The SN posteriors from the individual and joint analyses broadly overlap. In the full $w_0w_a$CDM analysis, the boundary-corrected marginalized posterior for the BAO scale peaks at $s_{\rm BAO}=1.00$, with a 95\% upper credible limit $s_{\rm BAO}<3.08$, using the same burn-in, weighting, and boundary-correction prescription as in Table~\ref{tab:baosys}. Because its prior is bounded below at unity, this result quantifies the additional covariance allowed by the data and should not be interpreted as evidence that the published BAO covariance is underestimated.

At the level of the nuisance parameters, adding external cosmological data mildly alters the inferred correlation structure between surveys. In particular, the correlation results remain similar after DESI DR2 BAO and Planck-lite are added. The (Pantheon+)$\times$(Union3) correlation changes from $0.44^{+0.16}_{-0.23}$ to $0.43^{+0.17}_{-0.22}$ and still shows the clearest positive value. The (Pantheon+)$\times$(DES--Dovekie) correlation changes from $0.20^{+0.24}_{-0.27}$ to $0.16^{+0.24}_{-0.26}$ and becomes weaker. The (Union3)$\times$(DES--Dovekie) correlation remains consistent with zero. Both analyses already allow the SN covariance scales to vary, so these changes describe the effect of adding external data rather than isolating the effect of covariance rescaling.

Overall, the cosmological constraints remain stable across the covariance models tested here, although the nuisance-parameter posteriors can change. This robustness test does not establish the absence of unmodeled systematic biases.

\subsection{Combination of BAO, CMB, SN data}
\label{sec:sn/results}
As a final step, we combine DESI DR2 BAO, Planck-lite, and the joint Pantheon+ + Union3 + DES--Dovekie likelihood. We compare the correlation-only model with the extended covariance model. In the extended model, the three SN correlation parameters, the three SN covariance-rescaling parameters, and $s_{\rm BAO}$ are allowed to vary.

The resulting constraints are shown in Fig.~\ref{fig:desi2snsyscor}. Table~\ref{tab:full_cosmology} reports the marginalized cosmological constraints for both cosmological models and both covariance prescriptions; the corresponding nuisance parameters are summarized in Table~\ref{tab:sn_nuisance_summary}. The three SN correlation coefficients vary in both prescriptions, so this comparison isolates the effect of freeing the covariance scales conditional on allowing inter-survey correlations.

\begin{figure}[t]
    \centering
    \includegraphics[width=\linewidth]{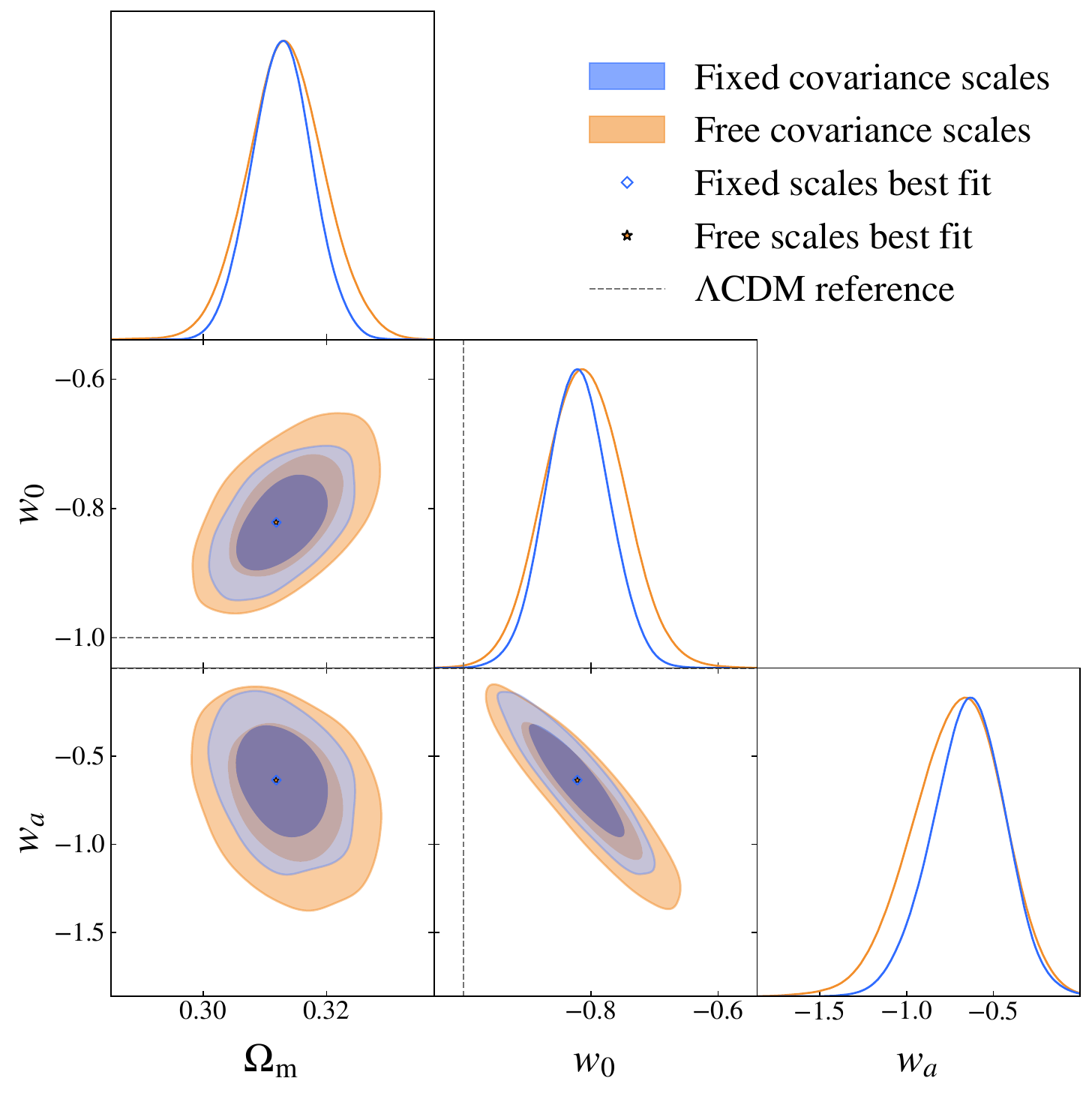}
    \caption{Marginalized constraints on $(\Omega_m,w_0,w_a)$ from DESI DR2 BAO + Planck-lite + joint SN under $w_0w_a$CDM. Blue and orange contours show fixed and free covariance scales, respectively, with SN correlations varied in both cases. Inner and outer contours enclose 68\% and 95\% of the marginalized posterior probability. The open blue diamonds and orange stars mark the corresponding best-fit points; dashed lines indicate $w_0=-1$ and $w_a=0$.}
    \label{fig:full_cosmology}
\end{figure}

\begin{table*}[t]
\centering
\small
\setlength{\tabcolsep}{12pt}
\renewcommand{\arraystretch}{1.25}
\begin{tabular}{@{}llccc@{}}
\hline\hline
Model & Covariance scales & $\Omega_m$ & $w_0$ & $w_a$ \\
\hline
$\Lambda$CDM & Fixed & $0.3029^{+0.0038}_{-0.0037}$ & $-1$ (fixed) & $0$ (fixed) \\
$\Lambda$CDM & Free & $0.3044^{+0.0055}_{-0.0047}$ & $-1$ (fixed) & $0$ (fixed) \\
$w_0w_a$CDM & Fixed & $0.3128^{+0.0048}_{-0.0048}$ & $-0.821^{+0.049}_{-0.048}$ & $-0.639^{+0.204}_{-0.218}$ \\
$w_0w_a$CDM & Free & $0.3133^{+0.0062}_{-0.0060}$ & $-0.811^{+0.063}_{-0.062}$ & $-0.698^{+0.241}_{-0.272}$ \\
\hline\hline
\end{tabular}
\caption{Marginalized cosmological constraints from Planck-lite + DESI DR2 BAO + joint Pantheon+ + Union3 + DES--Dovekie. Entries are weighted posterior medians with central 68\% credible intervals (16th--84th percentiles), after discarding the first 30\% of stored rows of each chain. These are posterior summaries, not optimization best-fit values. All three SN correlation coefficients vary in both prescriptions. Fixed scales means $s_{\rm PP}=s_{\rm U3}=s_{\rm DES}=s_{\rm BAO}=1$; free scales means that all four scales vary over their stated priors.}
\label{tab:full_cosmology}
\end{table*}

Freeing the covariance scales broadens the marginalized intervals while producing smaller changes in their medians. Under $w_0w_a$CDM, the median changes from $w_0=-0.821$ to $-0.811$ and from $w_a=-0.639$ to $-0.698$, while $\Omega_m$ changes from $0.3128$ to $0.3133$. Under $\Lambda$CDM, the median $\Omega_m$ changes from $0.3029$ to $0.3044$. The central 68\% intervals overlap between the fixed- and free-scale analyses. Since these analyses use the same observations, their posterior shifts are not independent measurements of a discrepancy.

Figure~\ref{fig:full_cosmology} complements Table~\ref{tab:full_cosmology} by showing the marginalized constraints on $(\Omega_m,w_0,w_a)$ for the full data combination under $w_0w_a$CDM. Allowing the covariance scales to vary broadens the credible regions, particularly along the $w_0$--$w_a$ degeneracy, while the fixed- and free-scale posteriors remain substantially overlapping. The posterior medians shift only mildly, and the two best-fit points nearly coincide.

\begin{table*}[t]
\centering
\footnotesize
\setlength{\tabcolsep}{7pt}
\renewcommand{\arraystretch}{1.18}
\begin{tabular}{@{}lcccc@{}}
\hline\hline
Parameter
&
\shortstack{Individual survey\\+ CMB + BAO\\$w_0w_a$CDM}
&
\shortstack{Joint SN (without CMB or BAO)\\$w_0w_a$CDM}
&
\shortstack{Full combination\\$\Lambda$CDM}
&
\shortstack{Full combination\\$w_0w_a$CDM}
\tabularnewline
\hline

$r_{\mathrm{PP,U3}}$
& $\cdots$
& $0.44^{+0.16}_{-0.23}$
& $0.50^{+0.15}_{-0.19}$
& $0.43^{+0.17}_{-0.22}$
\tabularnewline

$r_{\mathrm{PP,DES}}$
& $\cdots$
& $0.20^{+0.24}_{-0.27}$
& $0.32^{+0.20}_{-0.25}$
& $0.16^{+0.24}_{-0.26}$
\tabularnewline

$r_{\mathrm{U3,DES}}$
& $\cdots$
& $-0.10^{+0.27}_{-0.26}$
& $0.07^{+0.24}_{-0.25}$
& $-0.12^{+0.27}_{-0.24}$
\tabularnewline

\hline

$s_{\rm PP}$
& $1.30\;(<2.99)$
& $1.45\;(<3.11)$
& $1.41\;(<2.89)$
& $1.46\;(<3.04)$
\tabularnewline

$s_{\rm U3}$
& $1.00\;(<2.43)$
& $1.14\;(<2.63)$
& $1.46\;(<3.00)$
& $1.15\;(<2.67)$
\tabularnewline

$s_{\rm DES}$
& $1.00\;(<2.84)$
& $1.00\;(<2.76)$
& $1.35\;(<3.16)$
& $1.10\;(<2.87)$
\tabularnewline

\hline\hline
\end{tabular}

\caption{Constraints on the SN inter-survey correlation coefficients and covariance-rescaling parameters. For $r_{XY}$, we report posterior medians and central 68\% credible intervals. For $s_X$, we report the modes of the boundary-corrected one-dimensional marginalized posteriors, with 95\% upper credible limits in parentheses. The upper limits satisfy $\int_{1}^{s_{X,95}} p(s_X\mid d)\,{\rm d}s_X=0.95$. Statistics use chain weights after discarding the first 30\% of rows in each chain; the scale densities use first-order GetDist boundary correction and multiplicative bias correction. The individual-survey column combines each SN compilation separately with Planck-lite and DESI DR2 BAO under $w_0w_a$CDM, fixing $s_{\rm BAO}=1$ and including no inter-survey correlations. The joint-SN column uses the combined Pantheon+, Union3, and DES--Dovekie likelihood without CMB or BAO. The full-combination columns additionally include Planck-lite and DESI DR2 BAO. The covariance matrices are rescaled as $C'_X=s_X C_X$.}
\label{tab:sn_nuisance_summary}
\end{table*}

As summarized in Table~\ref{tab:sn_nuisance_summary}, the nuisance-parameter constraints remain broadly stable when Planck-lite and DESI DR2 BAO are added to the joint SN likelihood. The (Pantheon+)$\times$(Union3) correlation shows the clearest preference for a positive value, while the (Pantheon+)$\times$(DES--Dovekie) correlation is weaker and depends somewhat on the adopted cosmological model. The (Union3)$\times$(DES--Dovekie) correlation remains consistent with zero. The covariance-rescaling posteriors remain concentrated toward the lower part of the adopted one-sided prior and allow moderate additional covariance, but these results do not by themselves establish that the published covariance matrices are underestimated.

Allowing the covariance-rescaling parameters to vary only slightly broadens the cosmological contours and does not clearly shift the preferred region. The (Pantheon+)$\times$(Union3) correlation retains the clearest positive value, the (Pantheon+)$\times$(DES--Dovekie) correlation is weaker, and the (Union3)$\times$(DES--Dovekie) correlation remains consistent with zero. The $s_{\rm BAO}$ posterior is broad and has support near $s_{\rm BAO}=1$. Compared with the SN-only analysis, the three correlation coefficients change from $0.44$ to $0.43$, from $0.20$ to $0.16$, and from $-0.10$ to $-0.12$, respectively. These changes are small and do not show a general shift toward negative correlations.

Overall, these results indicate that while residual systematics and inter-survey correlations may influence the detailed nuisance parameter structure of the combined likelihood, they do not significantly alter the resulting cosmological constraints within the current observational precision.

Table~\ref{tab:bestfit_comparison} summarizes the best-fit comparisons associated with Figs.~\ref{fig:baosys}, \ref{fig:union3sys}, \ref{fig:wwa_sys_sn}, and \ref{fig:desi2snsyscor}. For the full joint-SN combination, allowing covariance rescaling lowers the best-found effective $\chi^2$ by \(0.93\) under \(\Lambda\mathrm{CDM}\) and by \(0.44\) under \(w_0w_a\mathrm{CDM}\). Consequently, the effective $\chi^2$ improvement obtained by extending \(\Lambda\mathrm{CDM}\) to \(w_0w_a\mathrm{CDM}\) decreases from \(12.81\) with fixed covariance scales to \(12.32\) with free scales, a reduction of \(0.49\). The corresponding nominal Gaussian-equivalent significance changes from \(3.15\sigma\) to \(3.07\sigma\). For the individual Union3 and Pantheon+ combinations, the relative fit improvement decreases from \(15.00\) to \(14.27\) and from \(7.13\) to \(6.70\), respectively. Across all available comparisons, freeing the covariance scales changes the best-found effective $\chi^2$ by less than one within either cosmological model. These results indicate that covariance-rescaling freedom only modestly changes the relative fit quality of the two cosmological models.
\begin{table*}[t]
\centering\small
\setlength{\tabcolsep}{5pt}
\renewcommand{\arraystretch}{1.15}
\begin{tabular}{@{}lccccc@{}}
\hline\hline 
\toprule
& \multicolumn{2}{c}{$\chi^2_{\Lambda}-\chi^2_{w_0w_a}$}
& \multicolumn{2}{c}{$\chi^2_{\mathrm{fixed}}-\chi^2_{\mathrm{free}}$}
& $(\chi^2_{\Lambda}-\chi^2_{w_0w_a})_{\mathrm{fixed}}
  -(\chi^2_{\Lambda}-\chi^2_{w_0w_a})_{\mathrm{free}}$
\\
\cmidrule(lr){2-3}\cmidrule(lr){4-5}
Dataset
& Fixed scales
& Free scales
& $\Lambda$CDM
& $w_0w_a$CDM
&
\\
\midrule
\hline BAO + CMB & 9.25 & 9.25 & 0.00 & 0.00 & 0.00 
\\ BAO + CMB + Union3 & 15.00 & 14.27 & 0.72 & 0.00 & 0.73 
\\ BAO + CMB + Pantheon+ & 7.13 & 6.70 & 0.47 & 0.04 & 0.43 
\\ BAO + CMB + DES--Dovekie & 11.36 & 10.66 & 0.70 & 0.00 & 0.70 
\\ BAO + CMB + joint SN & 12.81 & 12.32 & 0.93 & 0.44 & 0.49\\ \hline\hline
\end{tabular}
\caption{Differences in the best-found effective $\chi^2$ values. BAO and CMB denote DESI DR2 BAO and Planck-lite, respectively. In the column headings, \(\chi^2_{\Lambda}\) and \(\chi^2_{w_0w_a}\) denote the best-found effective $\chi^2$ under \(\Lambda\mathrm{CDM}\) and \(w_0w_a\mathrm{CDM}\). The first two numerical columns give the improvement obtained by extending \(\Lambda\mathrm{CDM}\) to \(w_0w_a\mathrm{CDM}\), separately for fixed and free covariance scales. The next two columns give the improvement obtained by freeing the covariance scales within each cosmological model. The final column gives the reduction in the relative best-fit advantage of \(w_0w_a\mathrm{CDM}\) after freeing the scales. For BAO + CMB, only \(s_{\mathrm{BAO}}\) is freed. In the individual-SN analyses, only the corresponding \(s_X\) is freed, with \(s_{\mathrm{BAO}}=1\) and no inter-survey correlations. In the joint-SN analysis, the three SN correlation coefficients vary in both prescriptions, while the three SN covariance scales and \(s_{\mathrm{BAO}}\) are either fixed to unity or varied together. The native likelihood terms, Planck calibration contribution, and covariance log-determinants are retained.}
\label{tab:bestfit_comparison}
\end{table*}

It is instructive to compare our results with the recent ``Unite'' analysis of Pantheon+ and DES-Dovekie by Camilleri et al.~\cite{Camilleri2026Unite}. Although both analyses combine contemporary Type Ia supernova data, they adopt fundamentally different approaches to combining the underlying surveys. The Unite analysis performs a joint reanalysis of Pantheon+ and DES-Dovekie, constructing a single internally homogenised supernova sample. This includes a common treatment of photometric calibration, light-curve fitting, intrinsic scatter, host-galaxy properties, selection, and bias corrections, thereby reducing differences between the original survey analyses before constructing the combined likelihood.

In contrast, we do not reprocess or homogenise the individual supernova compilations. Pantheon+ and DES--Dovekie are retained as separate datasets, with their published covariance matrices propagated into the joint likelihood. We instead account phenomenologically for possible residual inconsistencies between the surveys by allowing their covariance matrices to be rescaled and by introducing effective correlations between the datasets. Thus, while Unite combines the surveys by constructing a common supernova dataset at the data-analysis level, our approach combines distinct datasets at the likelihood level while explicitly allowing for uncertainty in their mutual covariance.

\begin{figure}
    \centering
    \includegraphics[width=\linewidth]{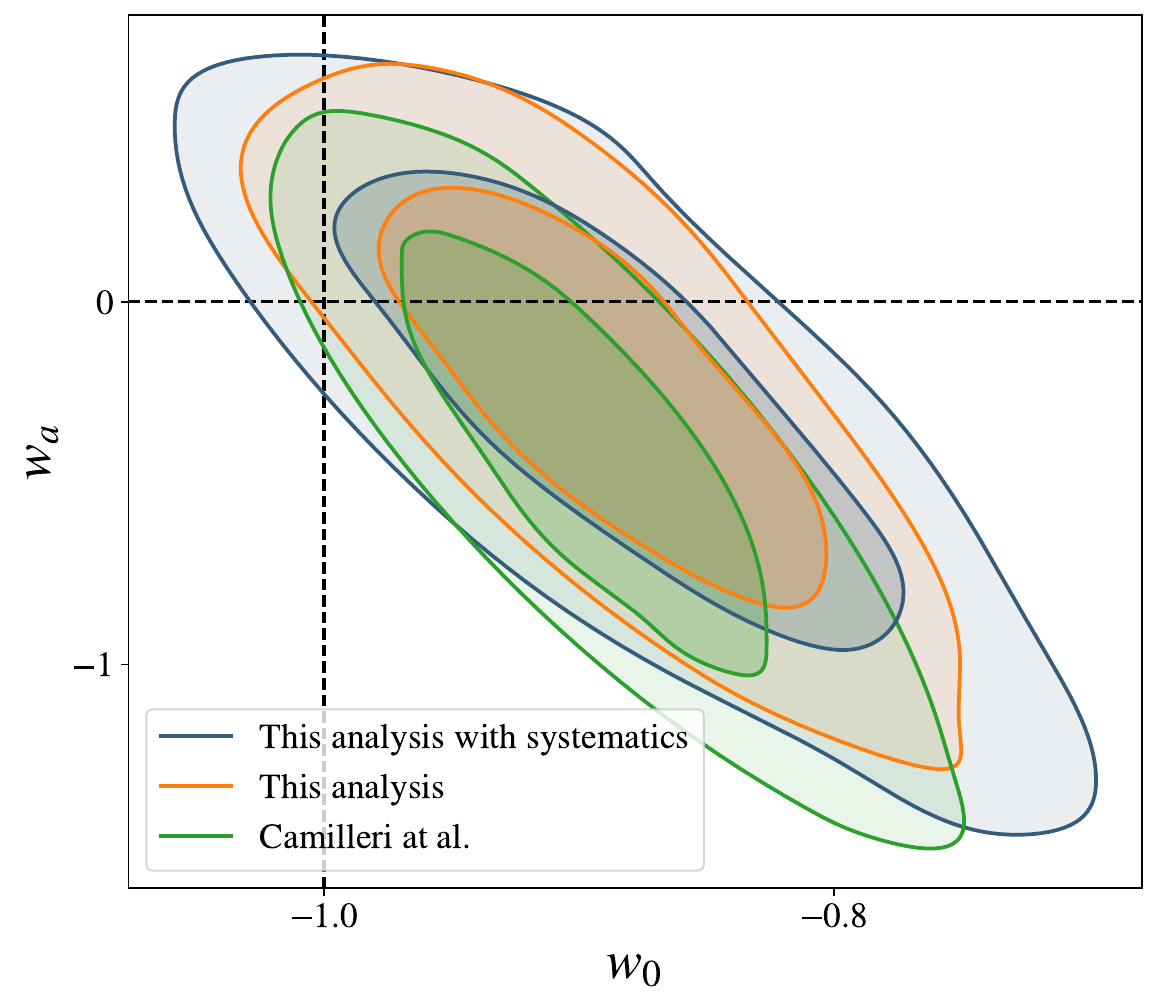}
    \caption{Constraints in the $w_0$--$w_a$ plane from Pantheon+ and DES-Dovekie combined with DESI DR2 BAO. Orange contours show our analysis with all covariance scales fixed to unity, and blue contours show the same analysis with free covariance-rescaling parameters; both vary the Pantheon+--DES correlation coefficient $r_{\rm PP,DES}$. Green contours show the ``Unite'' constraints~\cite{Camilleri2026Unite}, which combine the two surveys into a single homogenized sample with a joint covariance and include DES BAO, and DESI BAO data. Because Unite accounts for the cross-survey covariance but allows no additional covariance freedom, it is most directly comparable to the orange contours: the two show overlapping 68\%  and 95\% regions and a similar degeneracy direction, with our constraints broadly consistent. }
    \label{fig:comparison}
\end{figure}

For our Pantheon+ and DES-Dovekie combination with DESI DR2 BAO (but no CMB) using free inter-survey correlations, {\it without} systematics, the posterior medians and central 68\% intervals from Figure~\ref{fig:comparison} (orange contours) are
\begin{equation}
w_0 = -0.891^{+0.058}_{-0.053}, \qquad
w_a = -0.234^{+0.369}_{-0.406}.
\end{equation}
For comparison, allowing for additional systematics and inter-survey correlations to vary (blue contours in Figure~\ref{fig:comparison}) gives
\begin{equation}
w_0 = -0.884^{+0.072}_{-0.067}, \qquad
w_a = -0.282^{+0.422}_{-0.461}.
\end{equation}
The corresponding constraints from the Unite analysis \cite{Camilleri2026Unite} (green contours in Figure~\ref{fig:comparison}) are
\begin{equation}
w_0 = -0.890^{+0.057}_{-0.051}, \qquad
w_a = -0.47^{+0.41}_{-0.40}.
\end{equation}
Comparing our no-systematics result with Unite, the two analyses yield broadly compatible constraints in the $w_0$--$w_a$ plane, with overlapping $68/95\%$ credible regions and a similar degeneracy direction. Allowing the covariance scales and inter-survey correlations to vary shifts the posterior medians and changes the allowed parameter ranges. The overlap between our no-systematics constraints and those from Unite provides a qualitative consistency check between two different approaches to combining supernova datasets. However, it does not isolate the impact of covariance rescaling or independently validate the assumed cross-survey covariance, since the underlying supernova samples and their likelihood constructions differ.

\section{Conclusions}
\label{sec:conclusions}
In this work, we developed a phenomenological framework to investigate the impact of residual systematics and inter-survey correlations in combined cosmological analyses involving CMB, BAO, and Type Ia supernova data. Rather than attempting to model the physical origin of possible inconsistencies between datasets, we introduced effective nuisance parameters that rescale covariance matrices and encode correlations between surveys directly at the likelihood level. This approach provides a flexible and agnostic way to test the robustness of cosmological constraints against imperfectly modeled uncertainties.

We first applied this framework to DESI DR2 BAO combined with Planck-lite. Allowing $s_{\rm BAO}$ to vary only slightly broadens the cosmological contours. The $s_{\rm BAO}$ posterior is broad and has support near the baseline value. Since the prior only allows $s_{\rm BAO}\geq1$, this result does not show that the published DESI DR2 covariance is too small.

The SN covariance-rescaling posteriors are broad. For the full combination under $w_0w_a$CDM, the boundary-corrected marginalized posterior modes are $s_{\rm PP}=1.46$, $s_{\rm U3}=1.15$, and $s_{\rm DES}=1.10$, with respective 95\% upper credible limits of $3.04$, $2.67$, and $2.87$. Since $s_X$ multiplies the covariance, the standard deviations scale by $\sqrt{s_X}$. The priors only allow $s_X\geq1$, so these results quantify the additional covariance allowed by the data rather than demonstrating that the published covariance matrices are too small. Possible sources include calibration, selection effects, mixed survey samples, and differences in light-curve methods, but this analysis does not identify a specific source.

To consistently combine multiple SN compilations, we introduced a common redshift binning strategy that allowed all datasets to be projected onto a unified basis. This enabled the construction of a joint covariance matrix incorporating both survey-specific systematics and phenomenological inter-survey correlations. We checked that the compressed data vectors, covariance matrices, and theoretical predictions were calculated correctly. We also compared the raw Pantheon+ likelihood with the 22, 24, and 26 bins versions. The posterior regions mostly overlap, and the 24- and 26-bin results are closer to the raw likelihood than the 22-bin result. We therefore use this comparison as a robustness test and do not claim that the 22-bin compression is exactly lossless.

Across the covariance prescriptions considered here, the marginalized cosmological contours remain substantially overlapping. Allowing covariance rescaling and inter-survey correlations produces mild broadening along the existing parameter degeneracies, but no clear displacement of the preferred cosmological region is visible in the posterior contours. The additional nuisance parameters therefore provide a useful robustness test without materially changing the inferred expansion history at the current observational precision. The available minimizations show changes of less than one in the best-found effective $\chi^2$ when covariance rescaling is allowed. For the full joint-SN combination, the effective $\chi^2$ improvement obtained by extending \(\Lambda\mathrm{CDM}\) to \(w_0w_a\mathrm{CDM}\) decreases only modestly, from \(12.808\) with fixed covariance scales to \(12.322\) with free scales. The DESI, Union3, and DES--Dovekie data are consistent with their published covariances, while Pantheon+ shows a mild preference for additional covariance.

The three correlation parameters show a clear general pattern. The (Pantheon+)$\times$(Union3) correlation has the strongest preference for a positive value. The (Pantheon+)$\times$(DES--Dovekie) correlation is weaker and changes with the cosmological model, while the (Union3)$\times$(DES--Dovekie) correlation remains consistent with zero. This pattern does not clearly change after DESI DR2 BAO and Planck-lite are added. 

Overall, the phenomenological covariance extensions modify the nuisance-parameter posteriors but do not produce a clear displacement of the marginalized cosmological constraints within the models considered here. At the same time, the broad $s_X$ posteriors extend above one, but the cosmological constraints remain stable when this extra covariance freedom is included. Since the priors do not allow $s_X<1$, the results should be treated as a test of the covariance model rather than as direct evidence for missing physical systematics.

The framework presented here can be extended in several directions. A more realistic treatment of systematics could allow for redshift-dependent covariance modifications or structured forms of inter-survey correlations motivated by specific observational effects. In addition, future analyses could incorporate next-generation datasets with substantially improved statistical precision, where even small residual systematics may become cosmologically relevant. More generally, the phenomenological methodology developed in this work provides a simple diagnostic tool for assessing the robustness of combined cosmological analyses in the presence of imperfectly modeled uncertainties.

\begin{acknowledgments}
We thank Luohan Wang for collaboration in the initial stages of this project.  

This research was partially supported by the Natural Sciences and Engineering Research Council of Canada and Perimeter Institute for Theoretical Physics; research at the Perimeter Institute is supported in part by the Government of Canada through the Department of Innovation, Science and Economic Development Canada and by the Province of Ontario through the Ministry of Colleges and Universities. EK also thanks the Institute of Physics for their financial support through the Bell Burnell Graduate Scholarship Fund. We further acknowledge use of OpenAI ChatGPT and Anthropic Claude for feedback on manuscript and preparing the plotting scripts. 

\end{acknowledgments}
\appendix
\section{Binning strategy}
\label{app:binning}

\subsection{Common basis and theory projection}

The phenomenological cross-covariance model requires Pantheon+, Union3, and DES--Dovekie to share an index-aligned redshift basis. We adopt the native 22-point Union3 grid listed in Table~\ref{tab:union3-grid} \cite{Union3}; it defines only the bin geometry and the correspondence between surveys, not a common measurement at each node.

\begin{table}[htbp]
\centering
\caption{Union3 redshift nodes and bin edges defining the common 22-bin basis~\cite{Union3}.}
\label{tab:union3-grid}
\small
\setlength{\tabcolsep}{0pt}
\renewcommand{\arraystretch}{1.25}
\begin{tabular}{@{}p{0.96\columnwidth}@{}}
\hline
\raggedright $z_b^\star$: $0.05$, $0.10$, $0.15$, $0.20$, $0.25$, $0.30$, $0.35$, $0.40$, $0.45$, $0.50$, $0.55$, $0.60$, $0.65$, $0.70$, $0.75$, $0.80$, $0.8981$, $0.9963$, $1.0944$, $1.2322$, $1.3910$, $2.2623$ \tabularnewline
\hline
\raggedright $e_0,\ldots,e_{22}$: $0$, $0.075$, $0.125$, $0.175$, $0.225$, $0.275$, $0.325$, $0.375$, $0.425$, $0.475$, $0.525$, $0.575$, $0.625$, $0.675$, $0.725$, $0.775$, $0.84905$, $0.9472$, $1.04535$, $1.1633$, $1.3116$, $1.82665$, $3$ \tabularnewline
\hline
\end{tabular}
\end{table}

The internal boundaries are the midpoints of adjacent nodes,
\begin{equation}
e_b=\frac{z_b^\star+z_{b+1}^\star}{2},\qquad b=1,\ldots,21,
\label{eq:app-bin-edges}
\end{equation}
with $e_0=0$ and $e_{22}=3$. Bins are left-inclusive and right-exclusive, except for the final bin, which includes its upper endpoint.

The data vector, covariance, and theoretical prediction are projected as in Eq.~(\ref{eq:main-sn-projection}). Including the magnitude-offset nuisance parameter, the residual entering the common-grid likelihood is
\begin{equation}
\widetilde{\boldsymbol r}_X(\boldsymbol\theta)
=W_X\!\left[\boldsymbol d_X-\boldsymbol\mu_X^{\rm raw}(\boldsymbol\theta)
-\Delta M_X\boldsymbol 1_X\right].
\label{eq:app-residual-projection}
\end{equation}
For Pantheon+ and DES--Dovekie, $\boldsymbol\mu_X^{\rm raw}$ is evaluated at every original CMB-frame and heliocentric redshift before projection. We therefore do not approximate $W_X\boldsymbol\mu_X^{\rm raw}(\boldsymbol\theta)$ by evaluating the model at the nominal or effective bin redshifts. Union3 is already supplied on the adopted grid and is retained with $W_{\rm U3}=I_{22}$.

\subsection{Survey-specific construction}

Applying the strict Pantheon+ selection $z_{\rm HD}>0.01$ retains 1590 of the 1701 public entries \cite{PantheonPlusData}. We use the corrected apparent magnitude $m_{b,{\rm corr}}$ and the corresponding $1590\times1590$ statistical-plus-systematic covariance submatrix, without adding a separate $z_{\rm HD}$-error term. Objects are assigned to bins using their individual $z_{\rm HD}$ values.

Let $I_b$ be the set of Pantheon+ objects in bin $b$, $C_b\equiv(C_{\rm PP})_{I_bI_b}$, and $\boldsymbol 1_b$ a vector of $|I_b|$ ones. The local generalized least-squares (BLUE) weights are
\begin{equation}
\boldsymbol w_b=
\frac{C_b^{-1}\boldsymbol 1_b}
{\boldsymbol 1_b^{\mathsf T}C_b^{-1}\boldsymbol 1_b}.
\label{eq:app-blue-weights}
\end{equation}
They are evaluated with a Cholesky solve and inserted into row $b$ of $W_{\rm PP}\in\mathbb R^{22\times1590}$. The weights sum to unity in every bin, preserving a redshift-independent magnitude offset. Although they are determined from within-bin covariance blocks, the complete selected covariance is propagated through Eq.~(\ref{eq:main-sn-projection}), so correlations between different bins are retained.

The DES--Dovekie product uses the same nominal indexing, but only the first 19 bins, through $z^\star=1.0944$, contain observations \cite{DovekieCalibration,DESDovekie}. The rows at $z^\star=1.2322$, $1.3910$, and $2.2623$ are stored only to preserve the common indexing. With $P_{\rm DES}\in\mathbb R^{22\times19}$ denoting the embedding matrix for the populated bins, the active residual and covariance are
\begin{equation}
\boldsymbol r_{\rm DES}=P_{\rm DES}^{\mathsf T}
\widetilde{\boldsymbol r}_{\rm DES}^{(22)},\qquad
C_{\rm DES}^{(19)}=P_{\rm DES}^{\mathsf T}
\widetilde C_{\rm DES}^{(22)}P_{\rm DES}.
\label{eq:app-des-selection}
\end{equation}
The joint likelihood therefore contains $22_{\rm PP}+22_{\rm U3}+19_{\rm DES}=63$ observed entries. Empty DES--Dovekie bins are never treated as zero-valued measurements and do not enter the residual, covariance, or cross-covariance blocks.

\subsection{Compression validation}

\begin{figure*}[tp]
\centering
\includegraphics[width=0.7\textwidth]{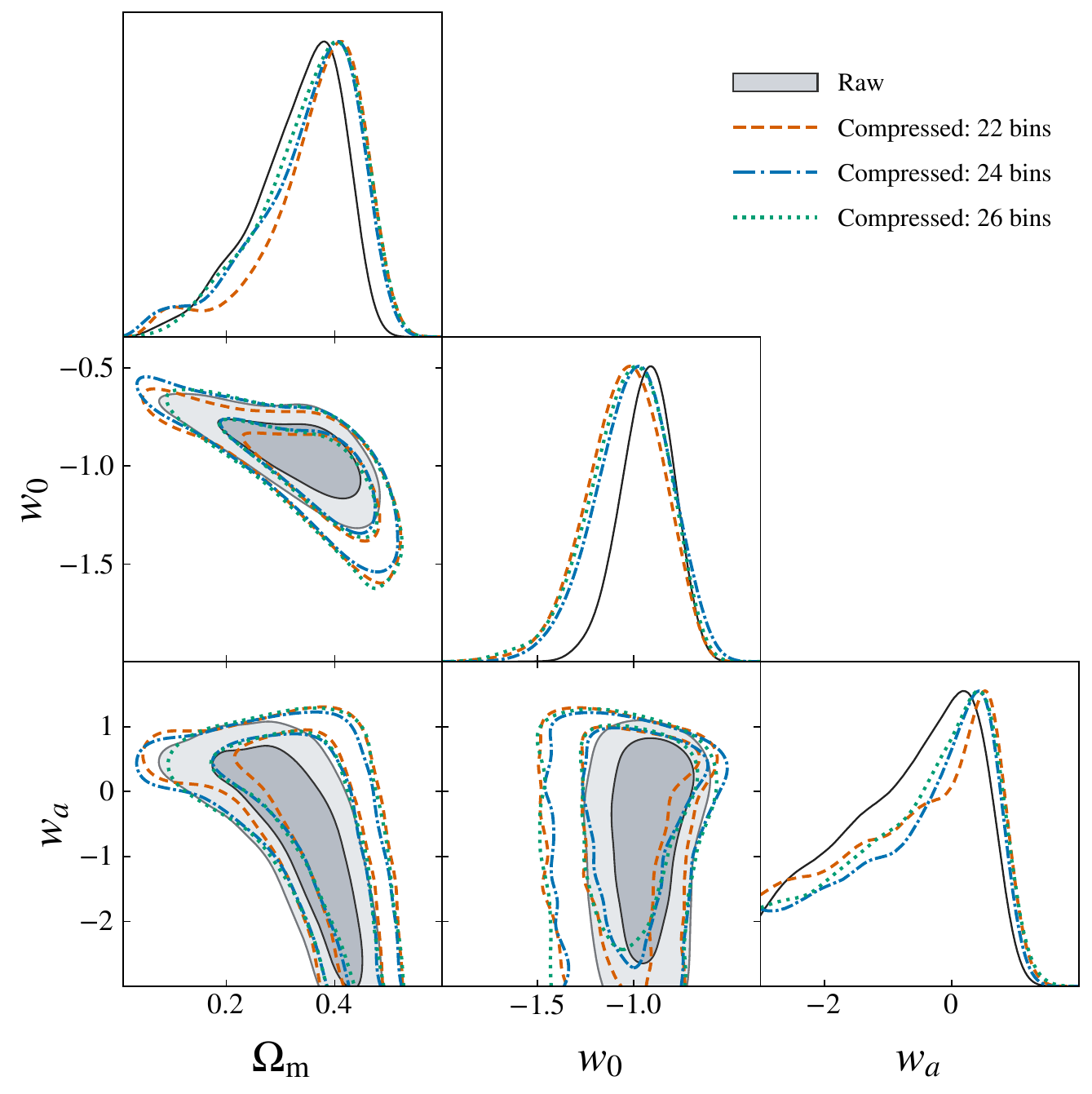}
\caption{Pantheon+ compression validation under $w_0w_a$CDM. Black contours show the raw 1590-supernova likelihood, while red, blue, and green contours show the 22, 24, and 26 bins results, respectively. Inner and outer contours enclose the 68\% and 95\% credible regions. The compression dependence is most visible along the weakly constrained $w_0$--$w_a$ directions, while all posterior regions remain substantially overlapping.}
\label{fig:pp-compression-w0wa}
\end{figure*}
Implementation checks reproduce the projected data and covariance to numerical precision, verify unit row sums for populated bins, and confirm that the active covariance matrices are symmetric and positive definite. These algebraic checks validate the implementation but do not establish that the compression preserves cosmological information.

We test cosmological robustness under $w_0w_a$CDM by comparing the raw Pantheon+ likelihood with the 22-, 24-, and 26-bin compressed likelihoods. These validation configurations contain only the Pantheon+ likelihood, with no Planck-lite CMB or BAO likelihood included. They share the same explicit cosmological parameter priors, including $-3\leq w_0\leq1$, $-3\leq w_a\leq2$, and a Gaussian prior $\omega_b=0.02218\pm0.00055$. The raw run samples the supernova magnitude offset directly, whereas the compressed runs use a re-parameterized offset. The 24- and 26-bin schemes share 18 approximately equal-count bins over $0.023\leq z_{\rm HD}<0.5$ and four fixed bins, $[0.5,0.6)$, $[0.6,0.7)$, $[0.7,0.85)$, and $[0.85,3]$; they use, respectively, two and four approximately equal-count bins over $0.01<z_{\rm HD}<0.023$. Boundaries are placed only between distinct redshifts.

As shown in Fig.~\ref{fig:pp-compression-w0wa}, the finer schemes generally provide comparable or improved agreement with the raw posterior, especially along the weakly constrained $w_0$--$w_a$ directions. The posterior regions nevertheless remain substantially overlapping, and the differences do not alter the inferred cosmological region within the marginalized uncertainties. We retain the 22-bin construction in the joint likelihood because it matches the native Union3 indexing; the comparison is therefore interpreted as an empirical robustness test, not as evidence that the projection is exactly lossless.

\section{Sampling and posterior summaries}
\label{app:sampling}
The four full-combination configurations use the Cobaya MCMC sampler~\cite{Cobaya} with proposal learning enabled, proposal scale $2.4$, learning interval $200$, oversampling power $0.4$, and oversampling thinning enabled. The configured stopping targets are $R-1<0.01$ for the parameter-mean diagnostic and $R-1<0.2$ for the 95\% confidence-limit diagnostic. The sampler-level burn-in setting is zero, while the posterior summaries in Table~\ref{tab:full_cosmology} discard the first 30\% of stored rows and retain the original sample weights.

The full-combination $w_0w_a$CDM runs impose uniform bounds $-3\leq w_0\leq1$ and $-3\leq w_a\leq0$; the $\Lambda$CDM runs fix $w_0=-1$ and $w_a=0$. Free SN covariance scales have uniform priors on $[1,5]$, and the free BAO scale has a uniform prior on $[1,9]$. Pairwise SN correlation coefficients have uniform bounds $[-1,1]$ subject to the positive-definiteness constraint in Eq.~\ref{eq:correlation-matrix}. A displayed BAO-scale range ending at 5 does not truncate or re-normalize its prior or posterior at 5.

The full-combination constraints reported here are conditional on the restricted range $-3\leq w_a\leq0$. The $\Lambda$CDM point lies on the upper boundary of this range, and the effect of extending the prior to positive $w_a$ has not been tested.

In the tables, cosmological parameters are summarized by weighted posterior medians and central 68\% credible intervals, defined by the weighted 16th and 84th percentiles. Covariance-scale parameters are instead reported as one-dimensional marginalized posterior modes with 95\% upper credible limits. The modes are estimated using GetDist with the hard prior ranges specified and first-order boundary and multiplicative bias corrections; the upper limits are weighted 95th percentiles of the retained samples. The two-dimensional contours enclose 68\% and 95\% of the marginalized posterior probability in highest-density regions. Best-fit markers denote numerical minima of the effective $\chi^2$, evaluated independently of posterior smoothing, and need not lie within the marginalized 68\% contours.

Best-fit values are obtained by numerically minimizing the effective $\chi^2$ with iminuit, subject to the parameter bounds of each analysis. We perform minimizations from multiple starting points and report the lowest effective $\chi^2$ found. These values are treated as best-found fits rather than certified global minima.

\section{Triangle plots}

\begin{figure*}[tp]
\centering
\includegraphics[width=0.9\textwidth,height=0.72\textheight,keepaspectratio]{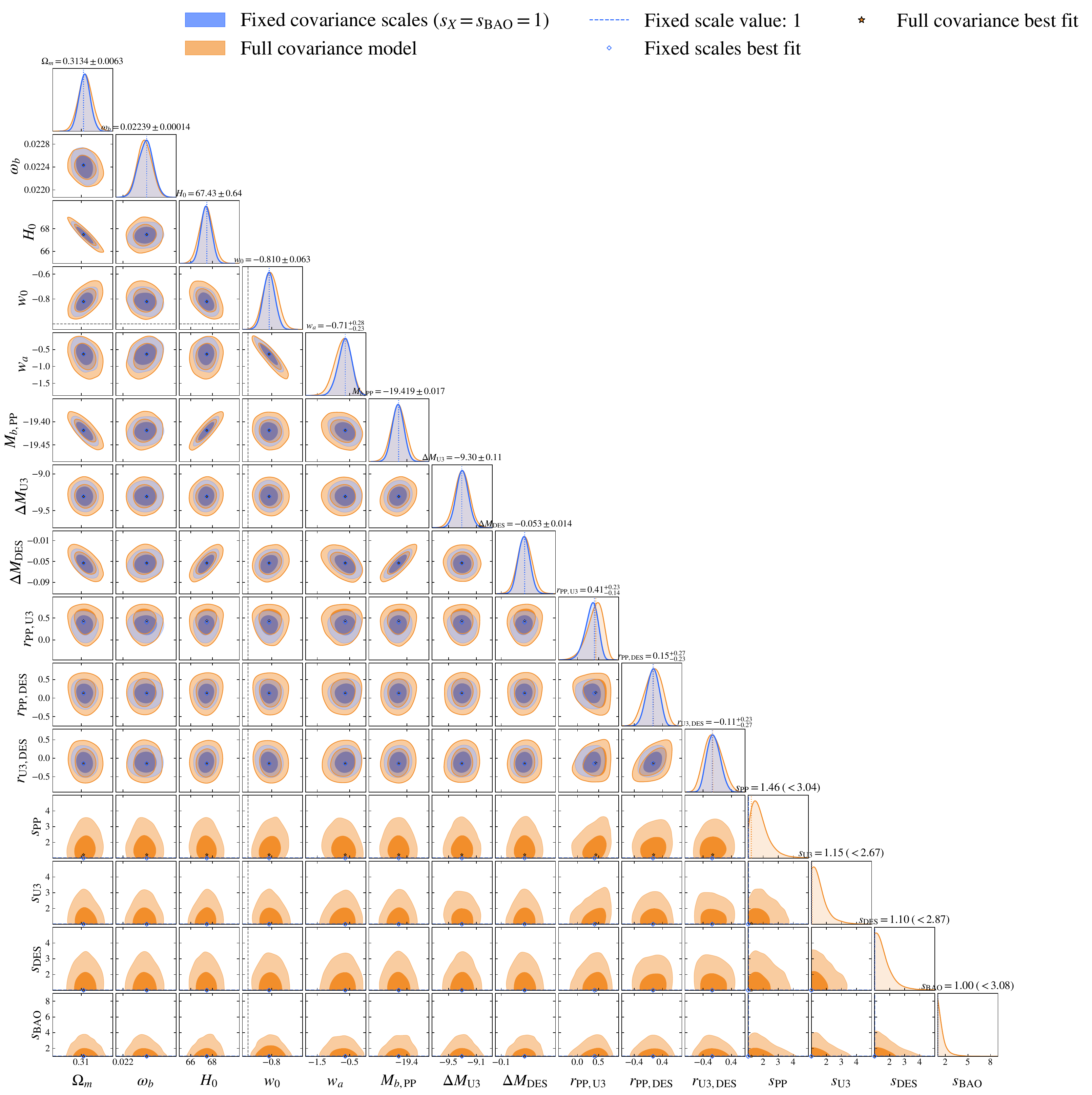}
\caption{Marginalized posterior distributions under $w_0w_a$CDM for the full combination of DESI DR2 BAO, Planck-lite, and the joint Pantheon+ + Union3 + DES--Dovekie likelihood. Only the 19 populated DES--Dovekie bins enter the likelihood. Blue contours show the correlation-only model with all covariance-rescaling parameters fixed to one, while orange contours show the model with free SN covariance-rescaling parameters and free $s_{\rm BAO}$. The contours show the 68\% and 95\% credible regions. The cosmological constraints remain almost unchanged. The (Pantheon+)$\times$(Union3) correlation shows the clearest positive value, the (Pantheon+)$\times$(DES--Dovekie) correlation is weaker, and the (Union3)$\times$(DES--Dovekie) correlation remains consistent with zero. The dashed lines mark the \(\Lambda\)CDM reference point, while the colored markers indicate the best-found effective-$\chi^2$ minima of the corresponding sampled models. All best-fit points lie within their respective 68\% credible contours.}
\label{fig:desi2snsyscor}
\end{figure*}

\begin{figure*}[tp]
    \centering
    \includegraphics[width=0.50\textwidth,height=0.26\textheight,keepaspectratio]{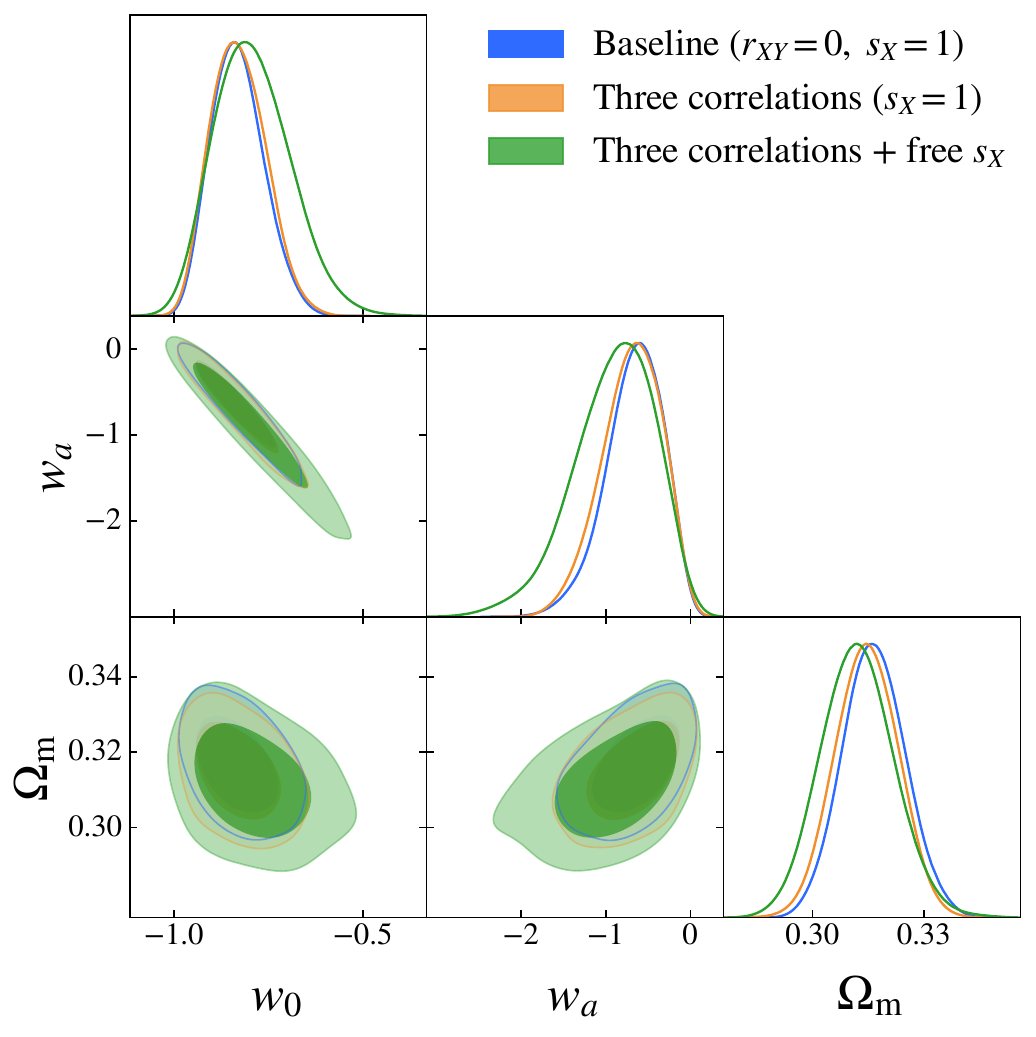}
    \par\vspace{0.4em}
    \includegraphics[width=0.68\textwidth,height=0.46\textheight,keepaspectratio]{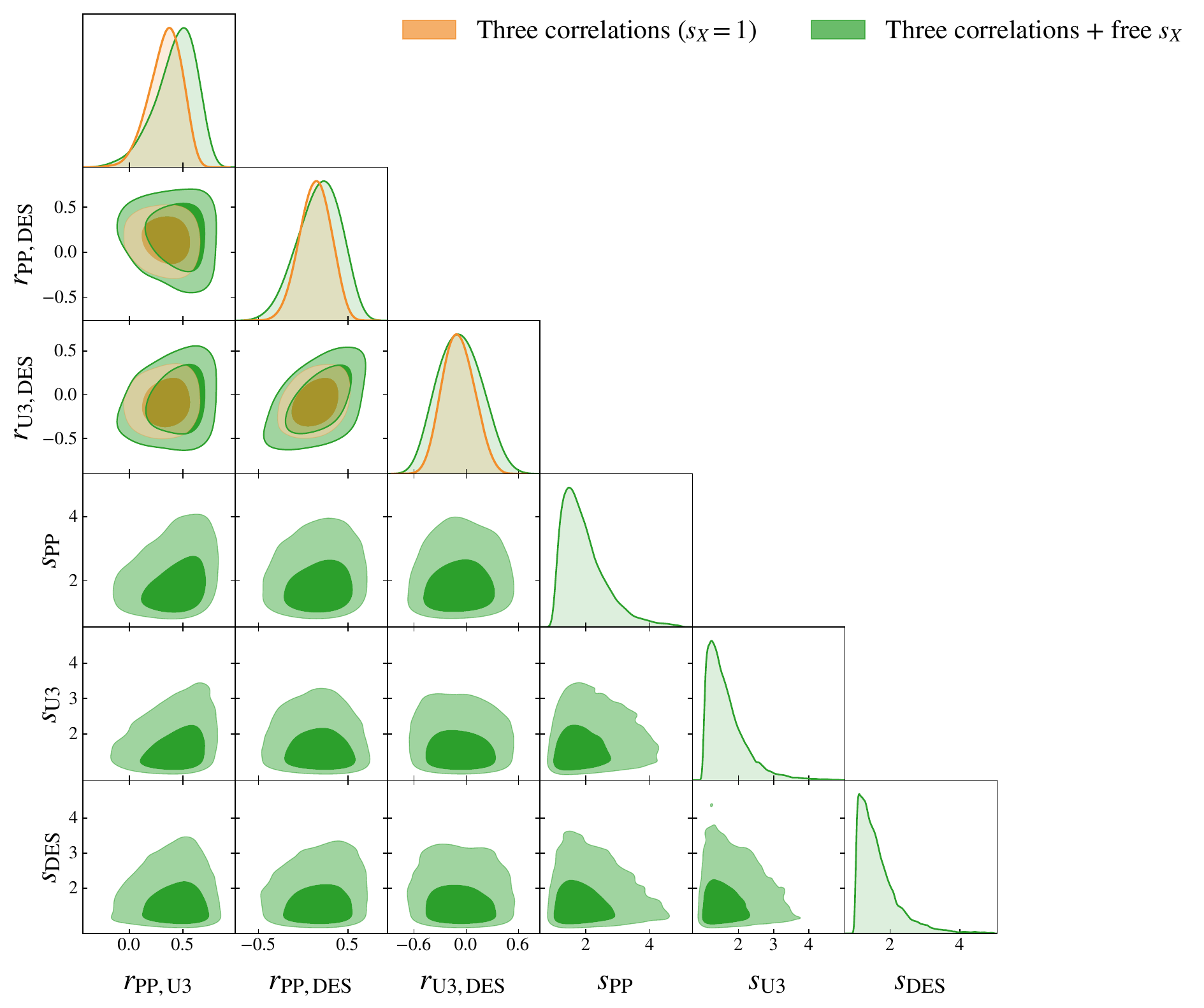}

    \caption{Final marginalized constraints from the combination of Planck-lite and the joint Pantheon+ + Union3 + DES--Dovekie likelihood under $w_0w_a$CDM. The upper panel shows the one- and two-dimensional marginalized posteriors for $(w_0,w_a,\Omega_m)$. Blue contours show the baseline model with $r_{ij}=0$ and all covariance-rescaling parameters fixed to unity. Orange contours show the correlation-only model, in which the three pairwise SN correlation coefficients are varied while all covariance scales remain fixed. Green contours show the full model, in which the three correlation coefficients and the three SN covariance-rescaling parameters are varied. The lower panel compares the SN nuisance-parameter posteriors from the correlation-only and full models. The filled contours enclose the 68\% and 95\% credible regions.}
    \label{fig:final_joint_constraints}
\end{figure*}
\clearpage
\bibliographystyle{apsrev4-2}
\bibliography{bibliography}

\end{document}